\documentclass[a4paper,nofootinbib,preprintnumbers,twocolumn,preprintnumbers,floatfix,superscriptaddress,pra,showpacs]{revtex4-1}

\usepackage{graphicx}
\usepackage{amssymb}
\usepackage{amsmath}
\usepackage{epstopdf}
\usepackage{graphics}
\usepackage[caption=false]{subfig}
\usepackage{float}
\usepackage{color}
\usepackage{hyperref}
\usepackage{bbold}

\allowdisplaybreaks[1]

\newcommand{\bra}[1]{\langle #1|}					
\newcommand{\ket}[1]{|#1\rangle}					

\newcommand{\abs}[1]{\left| #1 \right|} 
\newcommand{\avg}[1]{\langle #1 \rangle} 
\newcommand{\dibraket}[2]{\left< #1 \vphantom{#2} \right|\! \left. #2 \vphantom{#1} \right>} 

\newcommand{\figref}[1]{Fig.~\ref{#1}}

\DeclareMathAlphabet\mathbfcal{OMS}{cmsy}{b}{n}

\begin{document}

\title{Scalable photonic network architecture based on motional averaging in room temperature gas}

\author{J. Borregaard }
\affiliation{The Niels Bohr Institute, University of Copenhagen, Blegdamsvej 17, DK-2100 Copenhagen \O, Denmark}
\affiliation{Department of Physics, Harvard University, Cambridge, MA 02138, USA}
\author{M. Zugenmaier}
\affiliation{The Niels Bohr Institute, University of Copenhagen, Blegdamsvej 17, DK-2100 Copenhagen \O, Denmark}
\author{J. M. Petersen}
\affiliation{The Niels Bohr Institute, University of Copenhagen, Blegdamsvej 17, DK-2100 Copenhagen \O, Denmark}
\author{H. Shen}
\affiliation{The Niels Bohr Institute, University of Copenhagen, Blegdamsvej 17, DK-2100 Copenhagen \O, Denmark}
\author{G. Vasilakis}
\affiliation{The Niels Bohr Institute, University of Copenhagen, Blegdamsvej 17, DK-2100 Copenhagen \O, Denmark}
\author{K. Jensen}
\affiliation{The Niels Bohr Institute, University of Copenhagen, Blegdamsvej 17, DK-2100 Copenhagen \O, Denmark}
\author{E. S. Polzik}
\affiliation{The Niels Bohr Institute, University of Copenhagen, Blegdamsvej 17, DK-2100 Copenhagen \O, Denmark}
\author{A. S. S\o rensen}
\affiliation{The Niels Bohr Institute, University of Copenhagen, Blegdamsvej 17, DK-2100 Copenhagen \O, Denmark}

\date{\today}

\begin{abstract}
Quantum interfaces between photons and ensembles of atoms have emerged as powerful tools for quantum technologies. A major objective for such interfaces is high fidelity storage and retrieval of a photon in a collective quantum state of many atoms. This requires long-lived collective superposition states, which is typically achieved with immobilized atoms. Thermal atomic vapors, which present a simple and scalable resource, have, so far, only been used for continuous variable processing or for discrete variable processing on short time scales where atomic motion is negligible. We develop a theory based on the concept of motional averaging to enable room temperature discrete variable quantum memories and coherent single photon sources. We show that by choosing the interaction time so that atoms kept under spin protecting conditions can cross the light beam several times during the interaction combined with suitable spectral filtering, we erase the ``which atom'' information and obtain an efficient and homogenous coupling between all atoms and the light. Heralded single excitations can thus be created and stored as collective spinwaves, which can later be read out to produce coherent single photons in a scalable fashion. We demonstrate the feasibility of this approach to scalable quantum memories with a proof-of-principle experiment with room temperature atoms contained in microcells with spin protecting coating, placed inside an optical cavity. The experiment is performed at conditions corresponding to a few photons per pulse and clearly demonstrates a long coherence time of the forward scattered photons, which is the essential feature of the motional averaging.

\end{abstract}


\maketitle

Quantum systems can potentially enable powerful quantum computation \cite{ladd,feynman,shor} and highly secure quantum networks \cite{cirac,kimble,duan3,sangouard1}. Especially for the latter, it is essential that the information can be stored in quantum memories for processing~\cite{briegel,duan3}. To this end ensembles of cold atoms have previously been considered for quantum memories since their large number of atoms enables a strong light-atom interaction \cite{simon,chou,chen,gorshkov1}. Cold atoms, however, require extended cooling apparatus, which makes the scalability of such systems challenging. In contrast, room temperature atoms are much simpler to work with and have been used for a range of operations with continuous variables \cite{hammerer}. Several experiments with room temperature atomic cells have exploited a form of motional averaging where atoms move in and out of the beam many times during the interaction. By using a spin-preserving coating of the walls of the cell~\cite{hammerer}, the atoms can return back into the beam after colliding with a cell wall without losing the phase information. The averaging thereby removes the detrimental effect of atomic motion for continuous variable quantum information processing, where a specific optical mode is measured by homodyne detection~\cite{julsgaard,xiao,klein}. On the other hand, for discrete variables protocols based on the detection of photon clicks, the situation is different since photon counters select an almost instantaneous temporal mode. As a consequence, the efficiency of such systems for discrete variables where e.g. a single spin excitation is stored collectively in the ensemble is still limited by the incoherent atomic motion, which leaks ``which atom'' information and collapses the collective state~\cite{zhao2}.

Here, we introduce room temperature microcells as a new system for discrete variable, ensemble-based quantum information processing. We show theoretically how the detrimental effect of atomic motion can be circumvented in order to have an efficient and coherent interaction between an atomic ensemble and light at the single photon level. The proposed technique can be used to make efficient single photon sources with memories and offers a solution to scalable photonic networks based on room temperature atoms. The spin-protecting microcells investigated here can also be used for long distance quantum communication in DLCZ-like repeater protocols \cite{duan3,jiang,sangouard1} or quantum simulations~\cite{aspuru,aaronson,tillmann}. The scalability of the system compared to cryogenic or cold atoms systems opens up the possibility of employing large arrays of such systems combined with spatial multiplexing to enhance the communication rate~\cite{briegel, collins}. As opposed to previous ensemble-based experiments with discrete variables encoded in moving atoms~\cite{hosseini}, which typically rely on performing operations sufficiently fast that the atoms remain inside the laser beams, we show how a form of motional averaging similar to the one used for continous variable processing can be used to erase the ``which atom'' information. This novel approach to photon counting experiments alleviates the effect of atomic motion and can be seen as ``trapping" coherent spins in a box-like potential made up of the coated walls. We also present a proof-of-principle experiment demonstrating the effect.
Besides the specific experimental realization described here, the ideas behind this form of motional averaging are generally applicable and may be used in other systems where fluctuations of the coupling strength is an issue, such as ion crystals~\cite{herskind}. A related motional averaging of frequency fluctuations has previously been considered in super-conducting qubits~\cite{sorin}.

We consider a setup where an ensemble of atoms with a $\Lambda$-scheme level structure is kept in a small alkene coated cell~\cite{balabas,balabas2} (see \figref{fig:figure1}). A cell with quadratic cross section with side length $2L=150$ $\mu$m containing Cesium atoms was used in Ref.~\cite{vasilakis}, for which the average time between atom-wall collisions was $\sim$1.4 $\mu$sec and the coherence time was 10 ms. The atoms can thus endure several collision with the wall before loosing coherence making the cells suitable as quantum memories.
\begin{figure}
\centering
\includegraphics[width=0.5\textwidth]{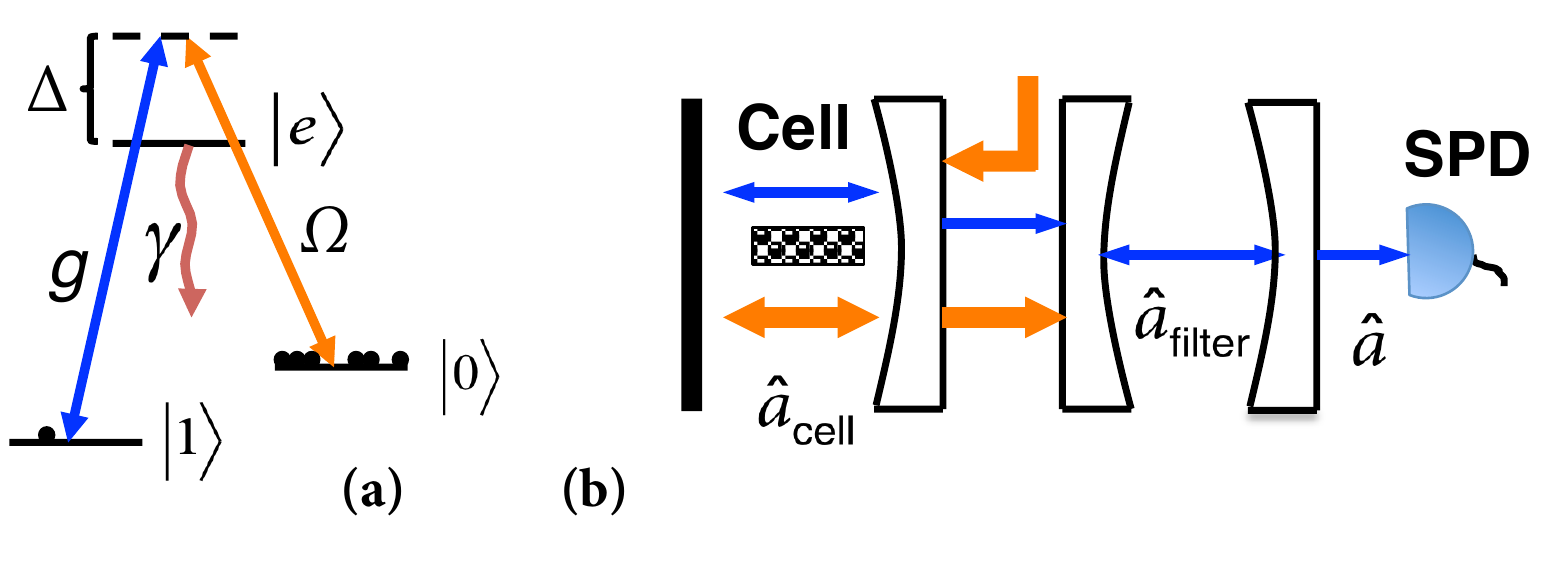}
\caption{Sketch of (a) the atomic level structure and (b) the experimental setup for creating a symmetric Dicke state. (a) All atoms are initially pumped to state $\ket{0}$. The transition $\ket{0}\to\ket{e}$ is driven by a weak laser field ($\Omega$) while the cavity mode ($g$) couples $\ket{e}$ and $\ket{1}$. (b) The atomic ensemble is kept in a small cell inside a single-sided cavity with a low finesse (cell-cavity). The quantum photons (thin arrows) are coupled from the cell-cavity into a high finesse cavity (filter-cavity), which separates them from the classical field (thick arrows) and averages over the atomic motion. Finally, the quantum photons are measured with a single photon detector (SPD).}
\label{fig:figure1}
\end{figure}
The ensemble is kept at room temperature and, to enhance the interaction with the light, the cell is placed inside a single-sided optical cavity (\emph{cell}-cavity). In the proof-of-principle experiment (see below) a finesse of $\mathcal{F}=17$ has been set by the output mirror transmission of 20\% and the reflection losses on the cell windows but a cavity with a higher finesse $\mathcal{F}=100$ can easily be envisioned. The light leaving the cell-cavity is coupled into another high finesse cavity (\emph{filter}-cavity), whose purpose is described below.

\begin{figure}
\centering
\includegraphics[width=0.45\textwidth]{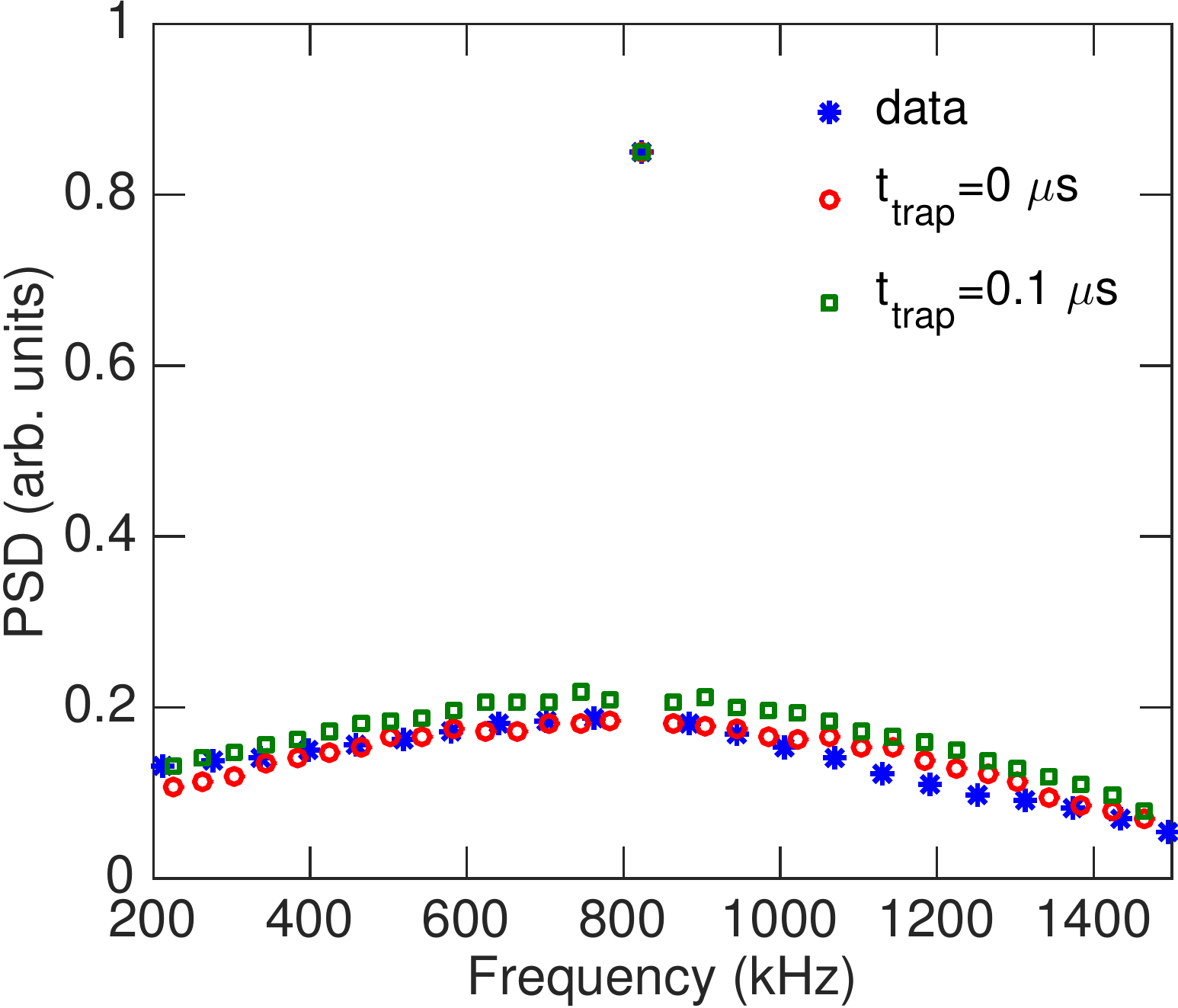}
\caption{Experimental data and simulation of the power spectral density (PSD) of the emitted light. The broad feature originates from short-time, incoherent light-atom interaction while the sharp peak is the long-time, coherent light-atom interaction. The optical scattering was obtained through the Faraday effect (see Sec.\ref{sec:experiment} for details) and is centered around the Larmor frequency at 823.8 kHz (the single high point in the figure). The simulations have been rescaled to coincide with the data at this point.  In the simulations we include the possibility of atoms being trapped in the  coating of the cell walls. From the figure, we estimate that such trapping time is below $0.1$ $\mu$s and can thus be ignored. The cell used in the experiment had dimensions $2L\times 2L\times 2L_{z}$ with $L=150$ $\mu$m and $L_{z}=0.5$ cm and the light beam had a Gaussian profile with a waist of $55$ $\mu$m. The statistical uncertainty of each experimental point is very small and is therefore not shown.}
\label{fig:figure2}
\end{figure}

Initially, all atoms are pumped to a stable ground state $\ket{0}$ (see \figref{fig:figure1}). In the \emph{write} process, the objective is to create a single, collective excitation in the ensemble, thereby creating the symmetric Dicke state $\ket{\psi_{\text{D}}}=\hat{S}_{\text{D}}\ket{00\ldots0}$, with $\hat{S}_{\text{D}}=\frac{1}{\sqrt{N}}\sum_{j}\ket{1}_{j}\bra{0}$ where $j$ is the atom number, $N$ is the total number of atoms and $\ket{1}$ is another stable ground state in the atoms. The $\ket{0}\to\ket{e}$ transition is driven with a laser pulse, which is far-detuned from the atomic transition to suppress the effect of the Doppler broadening of the atomic levels and absorption. In addition, the pulse should be sufficiently weak such that multiple excitations in the ensemble can be neglected. The write process is conditioned on detecting a single photon (quantum photon) emitted in a Raman transition $\ket{0}\to\ket{e}\to\ket{1}$. Upon detection, the atomic state is projected into the symmetric Dicke state if the light experienced a homogeneous interaction with all atoms, i.e. if the probability for different atoms to have emitted the photon is equal. In a realistic setup, the laser beam does not fill the entire cell and only atoms that are in the beam contribute to the cavity field, resulting in an asymmetric spin wave being created. Atoms leaving the beam will, however, return to the beam due to the frequent collisions with the cell walls. During the collisions, the atomic state is preserved due to the alkene coating of the cells and we exploit this to make a motional averaging of the light-atom interaction. If the interaction time is long enough to allow the atoms to move in and out of the beam several times, they will on average have experienced the same interaction with the light. Consequently, the detection of a cavity photon will, to a good approximation, project the atomic state to a Dicke state. Since the cell-cavity has a limited finesse, it may, in practice, not have a sufficiently narrow linewidth to allow this averaging. We therefore introduce an external filter-cavity. As we show below, the output from the cell-cavity consists of a spectrally narrow coherent component and a broad incoherent component (see \figref{fig:figure2}).
By selecting out the coherent part, the filter-cavity effectively increases the interaction time and allows for motional averaging. At the same time, the filter-cavity can also separate the quantum photon from the classical drive if there is a small frequency difference between the two such that only one frequency is resonant in the filter-cavity while both are sustained in the cell-cavity. Furthermore, choosing the difference to be an even number of free spectral ranges of the cell-cavity ensures an overlap of the field modes at the center of the cavity such that atoms can interact simultaneously with both modes if the length of the microcell is small compared to the wavelength corresponding to the frequency difference~\cite{SM}.     

After a successful creation of an excitation in the ensemble, the state can be kept until it is read out. In the readout process a long classical pulse addresses the $\ket{1}\to\ket{e}$ transition such that the single excitation is converted into a photon on the $\ket{e}\to\ket{0}$ transition (\figref{fig:figure1}a with $g$ and $\Omega$ interchanged).  This pulse should be long enough to allow for motional averaging as in the write process. The filter-cavity can once again be used to filter the quantum photon from the classical drive photons. Furthermore, it can also  be used to filter away incoherent photons as described below.

\section{Write process}
To characterize the quality of our system, when considered as a single photon source with memory, we first derive the efficiency of the write process and later discuss the readout efficiency and quality of the single photons being retrieved.
The Hamiltonian, describing the write process, is ($\hbar=1$)
\begin{equation} \label{eq:hamwrite}
\hat{H}=\sum_{j=1}^{N}\!-\!\Delta\hat{\sigma}_{ee}^{(j)}\!-\!\left(\frac{\Omega_{j}(t)}{2}\hat{\sigma}_{e0}^{(j)}\!+\!g_{j}(t)\hat{a}_{\text{cell}}\hat{\sigma}_{e1}^{(j)}\!+\!\text{H.c}.\right),
\end{equation}
where $\Delta=\omega_{\text{laser}}-\omega_{e0}$ with $\omega_{\text{laser}}$ being the frequency of the driving laser and $\omega_{e0}$ being the transition frequency between the levels $\ket{e}$ and $\ket{0}$. We have assumed that the cavity is on resonance with photons emitted on the $\ket{e}\to\ket{1}$  transition (see Fig.~\ref{fig:figure1}a). $\Omega_{j}$ ($g_{j}$) characterizes the coupling between the laser (cavity) field and the $j$'th atom. The field in the cell-cavity is described by the annihilation operator $\hat{a}_{\text{cell}}$ and we have defined the atomic operators $\sigma_{mn}^{(j)}=\ket{m}_{j}\bra{n}$ for the $j$'th atom, where $\{m,n\}\in\{0,1,e\}$. 

From the Hamiltonian, we obtain the following equations of motion
\begin{eqnarray}
\frac{d\hat{a}_{\text{cell}}}{dt}&=&-\frac{\kappa_{1}}{2}\hat{a}_{\text{cell}}+i\sum_{j=1}^{N}g^{*}_{j}(t)\hat{\sigma}_{1e}^{(j)} +\hat{F}_{\kappa_{1}} \label{eq:aeomt} \\~
\frac{d\hat{\sigma}_{1e}^{(j)}}{dt}&=&-\left(\frac{\gamma}{2}-i\Delta\right)\hat{\sigma}_{1e}^{(j)}-ig_{j}(t)\hat{a}_{\text{cell}}(\hat{\sigma}_{ee}^{(j)}-\hat{\sigma}_{11}^{(j)})\nonumber \\
&&+i\frac{\Omega_{j}(t)}{2}\hat{\sigma}_{10}^{(j)}+\hat{F}^{(j)}_{1e} \label{eq:eomt1} \\
\frac{d\hat{\sigma}_{10}^{(j)}}{dt}&=&-ig_{j}(t)\hat{a}_{\text{cell}}\hat{\sigma}_{e0}^{(j)}+i\frac{\Omega^{*}_{j}(t)}{2}\hat{\sigma}^{(j)}_{1e},
\end{eqnarray}
where we have included the cavity intensity decay with a rate $\kappa_{1}$ and the spontaneous emission of the atoms with a rate $\gamma$.  Associated with these decays, are corresponding Langevin noise operators, $\hat{F}_{\kappa_{1}}$, for the cavity decay and $\hat{F}_{1e}^{(j)}$ for the atomic decay \cite{gorshkov}. Note, that we have neglected dephasing of the atoms, e.g., due to collisions. We assume that all the atoms are initially in the ground state $\ket{0}$ and that the interaction with the light is a small perturbation to the system. We can therefore assume that $\hat{\sigma}^{(j)}_{ee}\approx\hat{\sigma}_{11}^{(j)}\approx0$. The noise operators describe vacuum noise and will never result in either an atomic or field excitation. Hence, they will never give rise to clicks in the detector (see \figref{fig:figure1}(b)) and we can consequently ignore them as described in Ref.~\cite{gorshkov}. Furthermore, we treat $\hat{\sigma}_{10}(t)$ as slowly varying in time and formally integrate Eqs.~\eqref{eq:aeomt} and \eqref{eq:eomt1} to obtain the field operator inside the cell
\begin{eqnarray}
\hat{a}_{\text{cell}}(t')&=&-\frac{1}{2}\sum_{j=1}^{N}\int_{0}^{t'}\text{d}t''\int_{0}^{t''}\text{d}t'''e^{-\kappa_{1}/2(t'-t'')}\nonumber \\
&&\times e^{-(\gamma/2-i\Delta)(t''-t''')}g^{*}_{j}(t'')\Omega_{j}(t''')\hat{\sigma}^{(j)}_{10}.
\end{eqnarray}

To find the field at the detector, we need to propagate the field through the filter-cavity. The input/output relations for the filter-cavity are
\begin{eqnarray}
\frac{d\hat{a}_{\text{filter}}}{dt}&=&-\frac{\kappa_{2}}{2}\hat{a}_{\text{filter}}+\sqrt{\kappa_{2}\kappa_{1}/2}\hat{a}_{\text{cell}} \label{eq:fileom}, \\
\hat{a}&=&\sqrt{\kappa_{2}/2}\hat{a}_{\text{filter}} \label{eq:outfil},
\end{eqnarray}
with $\kappa_{2}$ being the intensity decay rate of the filter-cavity, $\hat{a}_{\text{filter}}$ describes the field inside the filter-cavity and $\hat{a}$ describes the field at the detector. We have again neglected any input noise from the cavity decay since it never gives a click in our detector and we have also neglected intra-cavity losses. Formally integrating Eq.~\eqref{eq:fileom}, and using Eq.~\eqref{eq:outfil}, gives
\begin{equation}\label{eq:field}
\hat{a}=-\frac{\kappa_{2}\sqrt{\kappa_{1}}}{4}\sum_{j=1}^{N}\theta_{j}(t)\hat{\sigma}_{10}^{(j)},
\end{equation}
where
\begin{eqnarray} \label{eq:theta}
\theta_{j}(t)&=&\int_{0}^{t}\text{d}t'\int_{0}^{t'}\text{d}t''\int_{0}^{t''}\text{d}t'''e^{-\kappa_{2}(t-t')/2}e^{-\kappa_{1}(t'-t'')/2} \nonumber \\
&&\times e^{-(\gamma/2-i\Delta)(t''-t''')}g^{*}_{j}(t'')\Omega_{j}(t'''),
\end{eqnarray}

The efficiency is defined as the probability of having stored a single excitation in the symmetric Dicke state upon detection of a quantum photon. Neglecting higher order excitations, the atomic state is projected to $\sqrt{p(t)}\ket{\psi_{a}(t)}=\sqrt{\eta}~{}_{l}\!\bra{0}\hat{a}(t)\ket{00\ldots0}\ket{0}_{l}$ when the quantum photon is detected at time $t$. Here, $\ket{0}_{l}$ is the vacuum of the light in the cavity mode and $p(t)$ is the probability density of detecting the photon at time $t$ with $\eta$ being the single photon detection efficiency. Assuming that the driving pulse has a duration of $t_{\text{int}}$, the efficiency of the write process is
\begin{eqnarray} \label{eq:etawrite}
\eta_{\text{write}}\!=\!\frac{\int_{0}^{t_{\text{int}}}\!p(t)\!\abs{\dibraket{\psi_{\text{D}}}{\psi_{a}(t)}}^{2}\!\!\text{d}t}{\int_{0}^{t_{\text{int}}}p(t)\text{d}t}\!\approx\!\frac{\int_{0}^{t_{\text{int}}}\!\abs{\avg{\theta_{j}(t)}_{e}}^{2}\!\!\text{d}t}{\int_{0}^{t_{\text{int}}}\!\avg{\abs{\theta_{j}(t)}^{2}}_{e}\text{d}t},\!\qquad \!\quad
\end{eqnarray}
where we have used Eq.~\eqref{eq:field}, assumed $N\gg1$, and have defined the ensemble average $\avg{\ldots}_{e}=\frac{1}{N}\sum_{j=1}^{N}\avg{\ldots}$.

To get an expression for $\eta_{\text{write}}$, we need to evaluate $\abs{\avg{\theta_{j}(t)}}^{2}$ and $\avg{\abs{\theta_{j}(t)}^{2}}$. We now explicitly include the spatial dependence of the couplings assuming that $g_{j}(t)=g_{xy}^{(j)}(t)\sin(k_{q}z_{j}(t))$ and $\Omega_{j}(t)=\Omega_{xy}^{(j)}(t)\sin(k_{c}z_{j}(t))$, where
\begin{eqnarray} \label{eq:d6}
\Omega^{(j)}_{xy}(t)=\Omega e^{\frac{-x_{j}^{2}(t)-y_{j}^{2}(t)}{w^{2}}}, \\
g^{(j)}_{xy}(t)=g e^{\frac{-x_{j}^{2}(t)-y_{j}^{2}(t)}{w^{2}}},
\end{eqnarray}
with $w$ being the waist of the beams and ($x_{j}, y_{j},z_{j}$) is the position of the $j$'th atom. The transverse $xy$-dependence is assumed to be Gaussian while the $z$ dependence is sinusoidal due to the standing wave in the cavity.  $k_{q}$ ($k_{c}$) is the wave vector associated with the quantum photon (classical field). We have neglected additional geometric phases in the gaussian couplings since we are always considering the product $g_{j}^{*}\Omega_{j}$. 

The cavity field is a standing wave along the $z$-direction and both modes are assumed to have a node at the center of the cell at $z=0$. This geometry ensures an ideal overlap between the two modes at the position of the microcell at the center of the cavity. Away from the center of the cavity, the overlap of the two modes is degraded, which can lead to a detrimental phase difference if the length of the microcell along the cavity axis is to long. We estimate that for the Cs-cells used in the proof-of-principle experiment, the write efficiency is only degraded by a factor $\sim 0.97$ for a cell length of $\sim1$ cm~\cite{SM}. 

To suppress the effect of Doppler broadening of the atomic levels, $\Delta$ will be in the GHz range whereas the transverse waist of the beam will be on the order $50\mu$m. Consequently, the transverse coupling and the velocity of an atom can be considered constant for the integration over $t'''$ appearing in $\theta_{j}(t)$, which will have a typical time scale of $1/\abs{\Delta-i\gamma}$. The $z$-dependence of the coupling, however, varies rapidly due to the standing wave in the cavity and cannot be assumed to be constant. Writing $z_{j}(t''')=z_{j}(0)+v_{z}^{(j)}t'''$, where $v_{z}^{(j)}$ is the $z$-component of the velocity of the $j$'th atom, we can thus perform the integration over $t'''$ and adiabatically eliminate the optical coherence since we are far detuned. To obtain an expression for $\abs{\avg{\theta_{j}(t)}}^{2}$, we assume that the spatial distribution of the atoms is uniform and that the velocity distribution of the atoms follows the Maxwell-Boltzmann distribution with temperature $T$. Both distributions are assumed to be independent of time. In our analytical calculations, we also assume $k_{c}\sim k_{q}= k$ and that $kL_{z}\gg1$ such that $\avg{e^{\pm2ikz}}\approx0$, but in our numerical simulations presented below, we set the difference between  $k_{c}$ and $k_{q}$ corresponding to the real level structure of Cs where a splitting between $\ket{0}$ and $\ket{1}$ is 9.2 GHz. Here, $2L_{z}$ is the length of the cell in the beam direction.

With these assumptions, we obtain
\begin{eqnarray} \label{eq:thetax1}
\avg{\theta_{j}(t)}_{e}&=&\frac{\pi^{3/2}g\Omega}{4\Gamma_{d}}\mathbf{w}\left[(\Delta+i\gamma/2)/\Gamma_{d}\right]\frac{w^{2}}{L^{2}}\frac{1}{\kappa_{1}\kappa_{2}}, \label{eq:theta3}
\end{eqnarray}
where we have assumed that $e^{-(\kappa_{1}/2)t}\approx e^{-(\kappa_{2}/2)t} \approx 0$. Furthermore, we have assumed that the cell dimensions ($x\times y\times z$) are $2L\times2L\times2L_{z}$ and that $\text{erf}\left(\sqrt{2}L/w \right)^{2}\approx1$ meaning that we ignore any small portion of the beam, which is outside the cell. $\mathbf{w}[\ldots]$ is the Faddeeva function defined as $\mathbf{w}[z]=e^{-z^{2}}(1-\text{erf}(-iz))$ and $\Gamma_{d}=\sqrt{2k_{\text{B}}T/m}k$ is the Doppler width of the atomic levels at the temperature $T$ where $m$ is the atomic mass and $k_{\text{B}}$ is the Boltzmann constant.

While $\abs{\avg{\theta_{j}(t)}}^{2}$ does not contain any correlations between an atom's position at different times, $\avg{\abs{\theta_{j}(t)}^{2}}$ does. Eq.~\eqref{eq:etawrite} thus characterizes the effect of the random atomic motion and the motional averaging associated with it. We evaluate $\abs{\avg{\theta_{j}(t)}}^{2}$ under similar assumptions for the atoms as before. The correlations decay in time such that after many collisions with the walls, an atom's position is completely uncorrelated from its initial position. We assume that the decay of the correlations is exponential such that, e.g., $\avg{g^{(j)}_{xy}(0)g^{(j)}_{xy}(t)}=\avg{(g^{(j)}_{xy}(0))^{2}}e^{-\Gamma t}+\avg{g^{(j)}_{xy}(0)}^{2}(1-e^{-\Gamma t})$, where the first term contains the short time correlations while the second term characterizes the long time limit where the correlations are only through the average values. We substantiate this assumption by simulating a box of randomly moving, non-interacting atoms and find good agreement with a decay rate $\Gamma=\alpha v_{\text{thermal}}/w$ where $v_{\text{thermal}}$ is the average thermal velocity of the atoms, $w$ is the waist of the Gaussian cavity mode and $\alpha$ is a numerical constant on the order of unity~\cite{SM}. Employing this model for the atomic correlations and assuming $\kappa_{2}t_{\text{int}}\gg1$ such that the effective interaction time ($1/\kappa_{2}$) is set by the linewidth of the filter-cavity, we find that
\begin{eqnarray} \label{eq:thetax2}
\avg{\abs{\theta_{j}(t)}^{2}}&=&\abs{\avg{\theta_{j}(t)}_{e}}^{2}\left(1-2\kappa_{1}^{2}\kappa_{2}^{2}A(\kappa_{1},\kappa_{2},\Gamma)\right) \nonumber \\
&&+\abs{g}^{2}\abs{\Omega}^{2}A(\kappa_{1},\kappa_{2},\Gamma) \times \nonumber \\
&& \Bigg(\frac{\pi^{5/2}}{32\Gamma_{d}}\frac{w^{4}}{L^{4}}\Big(\frac{\text{Re}\left\{\mathbf{w}\left[(\Delta+i\gamma/2)/\Gamma_{d}\right]\right\}}{\gamma/2} \nonumber \\
&&+\frac{\text{Im}\left\{\mathbf{w}\left[(\Delta+i\gamma/2)/\Gamma_{d}\right]\right\}}{\Delta}\Big)\nonumber \\
&&+\frac{\pi^{2}}{4\Gamma_{d}^{2}}\abs{w\left[(\Delta+i\gamma/2)/\Gamma_{d}\right]}^{2}\frac{w^{2}}{L^{2}}\Bigg),
\end{eqnarray}
where we have defined
\begin{equation}
A(\kappa_{1},\kappa_{2},\Gamma)=\frac{(2\Gamma+\kappa_{1}+\kappa_{2})}{\kappa_{1}\kappa_{2}(2\Gamma+\kappa_{1})(2\Gamma+\kappa_{2})(\kappa_{1}+\kappa_{2})},
\end{equation}
and we have neglected all terms $\propto e^{2ikz_{j}}$ since these average to zero rapidly. Using Eqs. \eqref{eq:thetax1} and \eqref{eq:thetax2}, we can directly evaluate $\eta_{\text{write}}$ from Eq.~\eqref{eq:etawrite}. In the limit of $\kappa_{1}\gg(\Gamma,\kappa_{2})$ and $\Delta\gg\Gamma_{d}\gg\gamma$, the expression for $\eta_{\text{write}}$ reduces to
\begin{equation} \label{eq:writein}
\eta_{\text{write}}\approx\frac{1}{1+\frac{\kappa_{2}}{2\Gamma+\kappa_{2}}\left(\frac{4L^{2}}{\pi w^{2}}-1\right)} \approx1-\frac{1}{N_{\text{pass}}},
\end{equation}
where we have assumed $L > w$. Equation~\eqref{eq:writein} shows that $\eta_{\text{write}}\to1$ as $\kappa_{2}/\Gamma\to0$, i.e., the write efficiency improves with the length of the effective interaction time. This is the motional averaging of the atomic interaction with the light. Eq.~\eqref{eq:writein} also shows how the efficiency improves as the ratio between the beam area and the cell area ($\pi w^{2}/L^{2}$) increases. The last equality in Eq.~\eqref{eq:writein} is obtained assuming that $\Gamma>>\kappa_{2}$. In this limit $N_{\text{pass}}\approx\frac{\Gamma w^{2}}{\kappa_{2}L^{2}}$ can be interpreted as the average number of passes of an atom through the beam during the decay time of the filter-cavity.

\begin{figure}
\centering
\subfloat {\label{fig:figure3a}\includegraphics[width=0.42\textwidth]{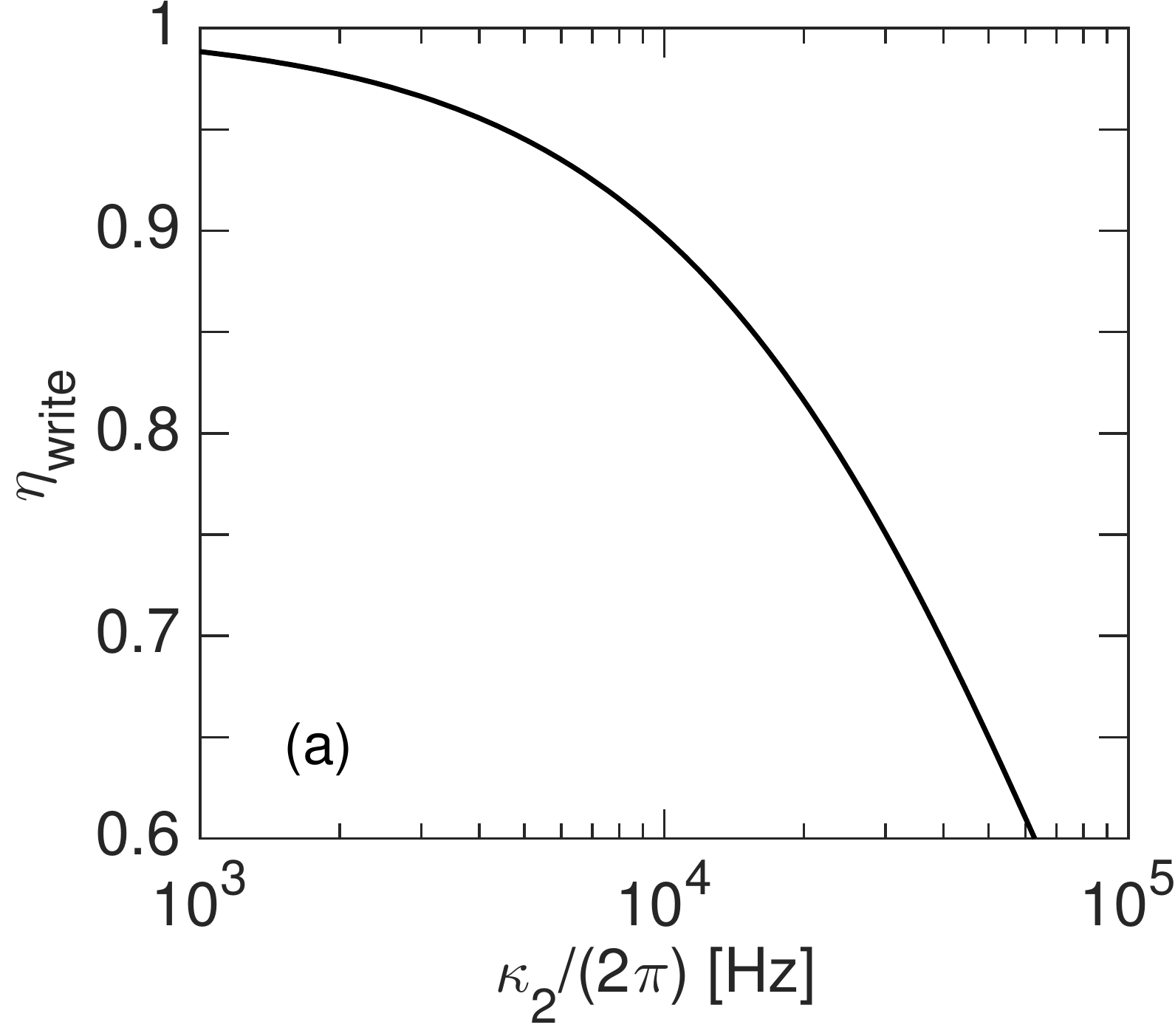}}\\
\subfloat{\label{fig:figure3b}\includegraphics[width=0.42\textwidth]{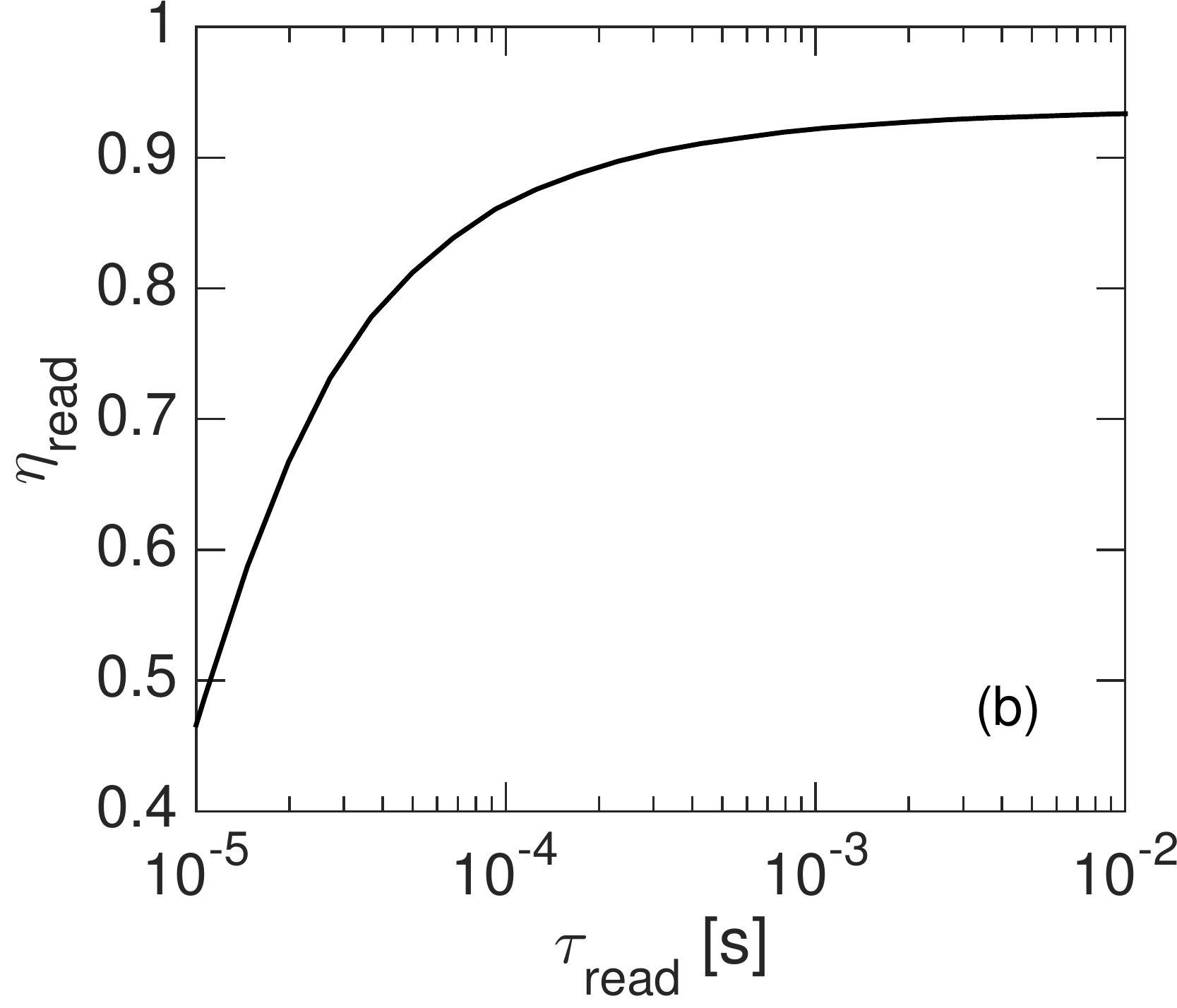}}
\caption{(a) Write efficiency as a function of the linewidth of the filter-cavity. We have simulated a Cs-cell with $L=150$ $\mu$m and $w=55$ $\mu$m corresponding to the cells being used in the proof-of-principle experiment. We have assumed a detuning of $\Delta\sim2\pi\cdot900$ MHz, a pulse length of $t_{\text{int}}=10/\kappa_{2}$ and $\kappa_{1}=2\pi\cdot46$ MHz. (b) Optimal readout efficiency as a function of the readout time $\tau_{\text{read}}$ without the filter cavity (corresponding to $\kappa_{2}\to\infty$). The efficiency was simulated for the same Cs-cells as the write efficiency and we have assumed that $\tau_{\text{read}}=3/\Gamma_{\text{read}}$ where $\Gamma_{\text{read}}$ is the readout rate, which is proportional to the classical drive intensity. The optical depth was assumed to be 168 as measured in the experiment. The finesse of the filter cavity was varied between 20 and 100 to get the optimal readout efficiency. We have included the full level structure of ${}^{133}$Cs in the simulations~\cite{SM}.}
\label{fig:figure3}
\end{figure}

To describe the write efficiency quantitatively, we have numerically simulated the experiment with Cs-cells including the full level structure of the atoms as described in Ref.~\cite{SM}. The $\Lambda$-scheme level structure can be realized with the two ground states $\ket{0}=\ket{F=4,m_F=4}$ and $\ket{1}=\ket{F=3,m_F=3}$ in the $6^{2}S_{1/2}$ the ground state manifold and the excited state $\ket{e}=\ket{F'=4,m_{F'}=4}$ in the excited $6^{2}P_{3/2}$ manifold. Note that with this configuration, the quantum and classical field differ both in polarisation and frequency and the filtering of the quantum photon is thus expected to be easily obtained using a combination of both  polarisation filtering and the filter-cavity. \figref{fig:figure3a} shows the simulated write efficiency as a function of $\kappa_{2}$.
It is seen that $\eta_{write}\approx90\%$ for $\kappa_{2}\approx2\pi \cdot 10$ kHz, which translates into a write time of $\sim150$ $\mu$s. Furthermore, we estimate that the number of classical photons, which should be filtered from the quantum photon is $\sim4.4\cdot10^{5}$ for realistic experimental parameters~\cite{SM}. This level of filtering is expected to be easily achieved using frequency filtering.  

\section{Proof-of-principle experiment} \label{sec:experiment}

To confirm the validity of the model and the results obtained above, we have performed a proof-of-principle experiment, which confirms the most important prediction, the presence of a spectrally narrow coherent peak of the scattered light arising from motional averaging. While several previous experiments~\cite{julsgaard,xiao,klein,balabas} have demonstrated long coherence times of room temperature atoms, here we wish to show a long coherence time of the emitted photons thus demonstrating that the motional averaging technique can be exploited to make coherent photon emission. To do this, we compare the theoretical predictions with the experimentally observed power spectral density (PSD) of light scattered by the atoms. In this proof-of-principle experiment, linearly polarized probe light, off-resonant from the atomic transition, interacts with the atoms resulting in  Faraday paramagnetic rotation of the light polarization \cite{hammerer} and the polarization state of the light is recorded with balanced polarimetry. As explained below, the balanced polarimetry establishes a heterodyne measurement of the Raman scattered photons allowing us to determine their spectrum.

\begin{figure}
\centering
\subfloat {\label{fig:CsLevelsa}\includegraphics[width=0.42\textwidth]{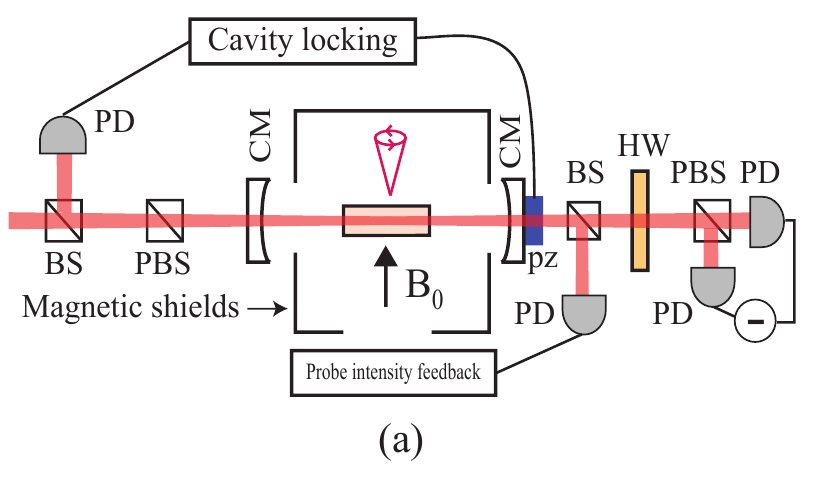}}\\
\subfloat{\label{fig:CsLevelsb}\includegraphics[width=0.42\textwidth]{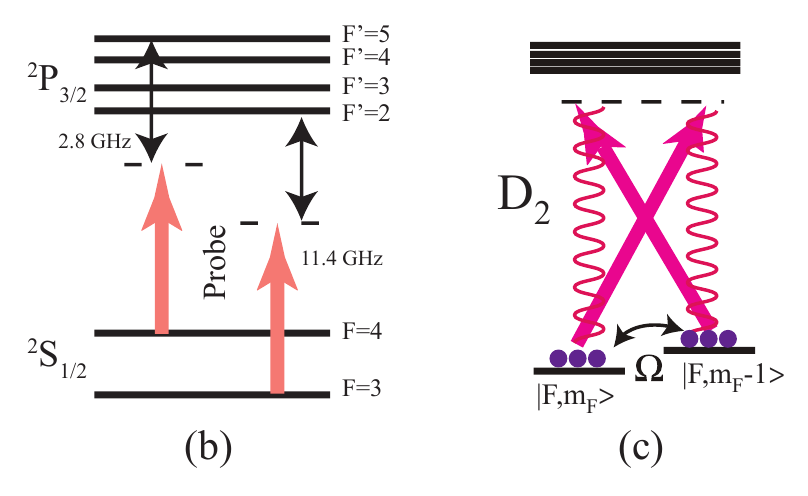}}
\caption{(a) Schematic of the proof-of-principle experiment. BS: Beam splitter; PBS: Polarizing Beam Splitter; CM: Cavity mirror; PD: Photodiode  (b) The D2 transition probed in the proof-of-principle experiment. (c) Effective coupling scheme for the Faraday interaction~\cite{hammerer}. The strong probe beam (straight arrows) is polarized perpendicular to the applied field and can thus drive $\sigma_+$ and $\sigma_-$ transition.   An atom scattered between two different $m_F$ levels results in a $\pi$ polarized photon (wiggly lines), orthogonal to the drive. In the weak probing limit, the measurement of the Faraday rotation angle is thus equivalent to a heterodyne measurement of the emitted light in the Raman transition with the probe pulse acting as a local oscillator. }
\label{fig:CsLevelsExp}
\end{figure}

The experimental setup is shown in Fig.~\ref{fig:CsLevelsExp}.
The light is red-detuned by 2.8 GHz from the $F=4 \mapsto F'=5 $ D2 transition.
For this detuning, the polarisation state of the probe light is affected by Cs atoms in both $F=4$ and $F=3$ ground state manifolds. 
The atomic ensemble is contained in a glass-cell with dimensions 300~$\mu$m$\times$300~$\mu$m$\times$1~cm corresponding to an average wall-to-wall time of flight of $\sim1.4$ $\mu$s. The walls of the cell are covered with an alkene coating \cite{balabas,balabas2}, resulting in longitudinal and transverse spin lifetime in the dark $T_1 \approx 17$~ms and $T_2 \approx 10$~ms, respectively. The atomic density inside the cell is estimated to be $\sim8\times 10^{-10}$~cm$^{-3}$ \cite{vasilakis}. The cell is placed inside a standing wave optical cavity to enhance the light-matter interaction. The cavity has finesse $\mathcal{F} \approx 17$, determined by the output coupler (intensity reflection $R_2 \approx 80 \%$) and the optical losses in the cell; currently the light intensity loss in the cell is $\sim 13\%$ per round trip, limited by the deterioration of the anti-reflection coating of the walls during the cell fabrication. A Pound-Drever-Hall technique is used to lock the cavity on resonance. The cavity mode has $\approx 55$~$\mu$m waist radius, which is a compromise between the requirement for strong coupling of light to the atomic ensemble and the requirement for low propagation losses through the cell.
A small portion of the beam at the cavity output is used in a feedback loop to compensate for the probe-intensity drift and maintain the same photon shot noise during the time of measurement.

A DC bias magnetic field perpendicular to the probe direction sets the Larmor frequency of the atoms. Due to technical limitations related with the phase noise of the laser and the cell birefringence, the polarisation of the probe was at an angle of approximately 40-45 degrees with respect to the axis of the magnetic field.  When the probe light is far detuned, the Faraday rotation  is, however, independent of this angle~\cite{hammerer}. For simplicity, we therefore describe the dynamics using the level structure in Fig.~\ref{fig:CsLevelsExp}(c), which assumes that the driving field  is $\sigma_+ +\sigma_-$ polarized, perpendicular to the direction of the magnetic field $\pi$. In the far detuned limit, the Faraday rotation is due to Raman transitions between magnetic states with magnetic quantum numbers $m_F$ differing by $\pm1$. In these Raman transitions, a $\pi$-polarized photon is emitted as shown in Fig.~\ref{fig:CsLevelsExp}(c).  In the balanced polarimetry, the driving field and the scattered $\pi$ component of the light are mixed on a polarising beam splitter and the difference intensity is recorded. This corresponds to the driving field acting as local oscillator for a heterodyne measurement of the emitted $\pi$-polarised light. The recorded spin noise is thus a measurement of the photons emitted from the atoms through Raman scattering between the Zeeman sublevels of the Cs hyperfine manifolds and is therefore exactly the quantity we are interested in for probing the coherence of the emitted photons and  verifying the predictions of the model. The experiment is performed in the continuous regime with constant laser intensity. By comparing the emitted light to the shot noise level, we can extract the Raman scattering rate. For a pulse duration that can lead to an efficient write step, e.g. $\tau=106$ $\mu$s, corresponding to $\kappa_2= 2 \pi \cdot 15$ kHz in \figref{fig:figure3}(a), we find that $\sim$8 Raman photons are scattered in the upper-sideband mode~\cite{SM}. Due to the linearity of the process, the spectrum is expected to be the same at the single photon level. 

The measurement is performed on atoms in approximately their thermal state i.e., the atoms are randomly distributed in the 16 m-sublevels of the F=3 and F=4 hyperfine manifold. There is a small deviation from the thermal state due to weak optical pumping from the probe, and all measurements are recorded in the resulting steady state. In this case, there is no macroscopic orientation and the probe induced back-action noise is negligible. The polarimetry noise is the sum of the photon shot noise and the Raman scattered photon noise.
The photon shot noise has a white power spectrum, whereas the spectrum of the recorded Raman noise is centered around the Larmor frequency due to the energy difference between magnetic sub-levels. 
We perform spin noise measurements for two different Larmor frequencies, $\sim 0.8$~MHz and $\sim 2.6$~MHz, at the same probe power. By subtracting the two power spectra, the photon shot noise and the electronic noise contribution to the recorded spectra can be removed and the Raman noise is acquired.

The measured PSD and the simulations of it are shown in \figref{fig:figure2}. The measured Raman noise reflects two different correlation decay timescales: a fast decay timescale $\sim$~1~$\mu$s associated with the transient time of flight through the probe beam, and a relatively slow decay $\sim100$~ $\mu$s, due to the spin decoherence (probe induced spin relaxation). Since the spectrum shown in \figref{fig:figure2} is measured for the scattered light, it is seen that the long spin coherence translates into a long coherence time of photons at the single photon level consistent with the theory. The PSD is recorded with a higher frequency resolution than shown in the figure but we bin the data with a frequency resolution ($\Delta f \approx 61$~kHz)  chosen so that the spin noise of atoms in the two hyperfine manifolds associated with the slow correlation decay timescale is contained in a single frequency bin. By doing this, complications arising from the nonlinear Zeeman splitting and the difference in the gyromagnetic ratio between the different hyperfine manifolds can be ignored.
The simulations were carried out as described in Ref.~\cite{SM} and have been rescaled to fit the measured SPD at f=823.7 kHz. We have carried out simulations both where the atomic collisions with the wall coatings happen instantaneously ($t_{\text{trap}}=0$), i.e. so that the trapping time is negligible compared to the transient time, and  with a trapping time of $t_{\text{trap}}=0.1$ $\mu$s. \figref{fig:figure2} shows an excellent agreement between the experiment and the model with zero trapping time. From \figref{fig:figure2}, we estimate that any trapping time in the experiment is $\lesssim0.1$ $\mu$s and can thus be ignored. The narrow peak in the scattered light is due to the fact that atoms repeatedly come back into the beam with the same spin phase, whereas the broad background is due to single transients through the beam. The narrow peak in the data thus demonstrates the long coherence time of the forward scattered light due to motional averaging. Considering the random motion of the atoms, the only coherence that can survive for that long is linked to the symmetric Dicke state as described in the theory. In essence, the idea of the motional averaging is to use a spectrally narrow filter cavity to select only the photons emitted in the narrow coherent peak. Since this narrow peak corresponds to a long interaction time, this means that all atoms participate equally in the resulting spin wave. Furthermore, since the narrow peak is much higher than the broad background, the loss in efficiency from the spectral filtering is limited. The excellent agreement between the simulation and the experiment thus confirms the applicability of the motional averaging as well as the theoretical model we use.

\section{Readout} \label{sec4}
We now consider the readout process. Assuming a single excitation has been stored in the symmetric mode in the ensemble, a classical drive ($\Omega$) is applied to read out the excitation as a cavity photon. The relevant Hamiltonian is obtained by interchanging $\hat{\sigma}_{e0}^{(j)}$ and $\hat{\sigma}_{e1}^{(j)}$ in Eq.~\eqref{eq:hamwrite}. From Heisenberg's equations of motion, we obtain
\begin{eqnarray}
\frac{d\hat{a}_{\text{cell}}}{dt}&=&-\frac{\kappa_{1}}{2}\hat{a}+i\sum_{j=1}^{N}g^{*}_{j}(t)\hat{\sigma}_{0e}^{(j)} \label{eq:aeom} \\
\frac{d\hat{\sigma}_{0e}^{(j)}}{dt}&=&-\left(\frac{\gamma}{2}-i\Delta\right)\hat{\sigma}_{0e}^{(j)}+ig_{j}(t)\hat{a}_{\text{cell}}+i\frac{\Omega_{j}(t)}{2}\hat{\sigma}_{01}^{(j)} \qquad \label{eq:eom1} \\
\frac{d\hat{\sigma}_{01}^{(j)}}{dt}&=&i\frac{\Omega^{*}_{j}(t)}{2}\hat{\sigma}^{(j)}_{0e},
\end{eqnarray}
where we have assumed that $\hat{\sigma}_{ee}^{(j)}-\hat{\sigma}_{00}^{(j)}\approx-\mathbb{1}$ and that the dynamics of $\hat{\sigma}_{10}^{(j)}$ are governed by the classical drive ($\Omega$). Furthermore, we have neglected the noise operators associated with spontaneous emission and cavity decay, as in the write process. We can formally integrate Eq.~\eqref{eq:eom1}, assuming the $xy$-dependence of the couplings to be constant for the integration while the $z$-dependent parts are of the form $\sin(k(z_{j}(0)+v_{z}^{(j)}(0)t))$, as in the write process. The integration gives a set of coupled equations
\begin{eqnarray}
\frac{d\hat{a}_{\text{cell}}}{dt}&=&\mathcal{A}(t)\hat{a}_{\text{cell}}+\sum_{j=1}^{N}\mathcal{B}_{j}(t)\hat{\sigma}_{01}^{(j)} \label{eq:supsys1}\\
\frac{d\hat{\sigma}_{01}^{(j)}}{dt}&=&\mathcal{B}_{j}(t)\hat{a}_{\text{cell}}+\mathcal{C}_{j}(t)\hat{\sigma}_{01}^{(j)}, \label{eq:supsys2}
\end{eqnarray}
where
\begin{eqnarray}
\mathcal{A}(t)&=&\frac{-\kappa_{1}}{2}+\frac{1}{4}\sum_{j=1}^{N}\abs{g_{xy}^{(j)}(t)}^{2}f_{j}(\gamma,\Delta,k) \label{eq:suppa1}\\
\mathcal{B}_{j}(t)&=&\frac{1}{8}(g_{xy}^{(j)}(t))^{*}\Omega_{xy}^{(j)}(t)f_{j}(\gamma,\Delta,k) \label{eq:suppbj} \\
\mathcal{C}_{j}(t)&=&\frac{1}{16}\abs{\Omega_{xy}^{(j)}(t)}^{2}f_{j}(\gamma,\Delta,k), \label{eq:suppcj} \\
f_{j}(\gamma,\Delta,k)&=&\Bigg(\frac{e^{2ikz_{j}(t)}-1}{\gamma/2-i(\Delta-kv_{z}^{(j)}(t))}\nonumber \\
&&-\frac{1-e^{-2ikz_{j}(t)}}{\gamma/2-i(\Delta+kv_{z}^{(j)}(t))}\Bigg),
\end{eqnarray}
and we have assumed that $k_{c}\approx k_{q} \approx k$. The equations can be expressed as a matrix system of the form
\begin{equation}
\frac{\text{d}\mathbf{x}(t)}{\text{d}t}=\mathbf{M}(t)\mathbf{x}(t),
\end{equation}
where $\mathbf{x}(t)=\left(\hat{a}_{\text{cell}},\hat{\sigma}_{01}^{(1)},\hat{\sigma}_{01}^{(2)},\ldots,\hat{\sigma}_{01}^{(N)}\right)$ and $\mathbf{M}(t)$ is the coupling matrix between the atoms and the cavity field. We now split the couplings into (large) average time-independent parts, $\{\avg{\mathcal{A}}_e,\avg{\mathcal{B}}_e,\avg{\mathcal{C}}_e\}=\{\bar{\mathcal{A}},\bar{\mathcal{B}},\bar{\mathcal{C}}\}$, and (small) time-dependent parts, $\{\delta\mathcal{A}(t),\delta\mathcal{B}_{j}(t),\delta\mathcal{C}_{j}(t)\}$. The coupling matrix can then be expressed as $\mathbf{M}(t)=\mathbf{M}_{0}+\delta\mathbf{M}(t)$ where $\mathbf{M}_{0}$ contains the average time-independent couplings while $\delta\mathbf{M}(t)$ contains the time dependent fluctuations. Assuming the readout pulse is long, the atoms will have had the same average interaction with the light meaning that $\mathbf{M}(t)\approx \mathbf{M}_{0}$.  Treating $\delta\mathbf{M}(t)$ as a small perturbation, we can then obtain a perturbative expansion of $\hat{a}_{\text{cell}}$. Assuming that the initial state of the atoms before readout is the symmetric Dicke state, we find that, to second order in $\delta\mathbf{M}(t)$, the cavity field can be expressed as $\hat{a}_{\text{cell}}\sim \hat{a}_{\text{cell}}^{(0)}+\hat{a}_{\text{cell}}^{(2)}$. Here we have omitted the first order term, which depends on the fluctuations in $\delta\mathcal{A}(t)$, since we find that its contribution is suppressed by a factor of at least $d\mathcal{F}/N$ compared to the other terms. Here $d/\tau_{\text{cav}}\propto Ng^{2}/\gamma$ is the optical depth on the $\ket{0}\leftrightarrow\ket{e}$ transition per cavity roundtrip time and $\mathcal{F}=2\pi/(\tau_{\text{cav}}\kappa_{1})$ is the finesse of the cell-cavity.

As in the write process, the field from the cell-cavity is sent through the filter-cavity in order to both filter the classical drive photons from the single photon and to filter out incoherent photons as we will describe below. Using Eqs.~\eqref{eq:fileom}-\eqref{eq:outfil}, we find that the readout efficiency is
\begin{eqnarray}
\eta_{\text{read}}&=&\frac{\kappa_{2}^{2}\kappa_{1}}{4}\int_{0}^{\tau_{\text{read}}}\!\!\!\!\!\!\!\!\!dt\int_{0}^{t}\!\!\!dt'\!\!\!\int_{0}^{t}\!\!\!dt''e^{-\kappa_{2}/2(2t-t'-t'')}\nonumber \\
&&\times\avg{\hat{a}_{\text{cell}}^{\dagger}(t')\hat{a}_{\text{cell}}(t'')}, \label{eq:etar}
\end{eqnarray}
where $\tau_{\text{read}}$ is the duration of the readout stage.
To lowest order we find
\begin{eqnarray}
a_{\text{cell}}(t)\approx\hat{a}_{\text{cell}}^{(0)}=\frac{\sqrt{N}\bar{\mathcal{B}}}{\sqrt{\mathcal{D}}}e^{\frac{1}{2}(\bar{\mathcal{A}}+\bar{\mathcal{C}}+\sqrt{\mathcal{D}})t}\left(1-e^{-\sqrt{\mathcal{D}}t}\right), \qquad \label{eq:eta0sup}
\end{eqnarray}
where $\mathcal{D}=(\bar{\mathcal{C}}-\bar{\mathcal{A}})^{2}+4N\bar{\mathcal{B}}^{2}$. Inserting Eq.~\eqref{eq:eta0sup} in Eq.~\eqref{eq:etar} and taking the limit of $\Omega\to0$ and $\tau_{\text{read}},\kappa_{2}\to\infty$, gives a zeroth-order readout efficiency of
\begin{equation} \label{eq:readeff}
 \eta_{\text{read},0}\approx\frac{1}{\frac{\pi}{d\mathcal{F}}+1}.
\end{equation}
Eq.~\eqref{eq:readeff} is equivalent to the result for cold atomic ensembles \cite{gorshkov} and represents the long time limit of perfect motional averaging where the efficiency improves with optical depth and finesse of the system.

The coherence time of real atoms is, however, limited and a fast readout is therefore desirable. From Eq.~\eqref{eq:eta0sup}, we identify the readout rate $\Gamma_{\text{read}}=\frac{1}{2}(\bar{\mathcal{A}}+\bar{\mathcal{C}}+\sqrt{\mathcal{D}})$. This readout rate $\Gamma_{\text{read}}$ increases with increasing strength of the readout pulse, so that for strong driving corresponding to a fast readout, it is necessary to consider higher order terms in the perturbative expansion of $\hat{a}_{\text{cell}}(t)$. To second order, we find that $\eta_{\text{read}}\sim \eta_{\text{read},0}+\eta_{\text{read},2}$ where the second order term ($\eta_{\text{read},2}$) mainly describes the loss of the excitation due to spontaneous emission. Consequently, the magnitude of $\eta_{\text{read},2}$ increases with the driving strength while it's sign is negative. $\eta_{\text{read},2}$ contains correlations between an atom's position at different times, which we can treat in a similar manner as in the write process, i.e. as exponentially decaying in time. By simulating the readout process with the Cs-cells in a similar fashion as for the write process, we can quantitatively describe the readout efficiency to second order~\cite{SM}. \figref{fig:figure3b} shows the readout efficiency to second order 
as a function of the readout time. We have assumed an optical depth of 168 and varied the finesse of the cell-cavity between 20-100 to get the maximum readout efficiency. The readout time is set to $\tau_{\text{read}}=3/\Gamma_{\text{read}}$ ensuring that a negligible population is left in the system at the end of the read out stage.  (Note that in these simulation we do not include the filter cavity considered for the write stage. Formally this corresponds to taking the $\kappa_{2}\to\infty$). The full level structure of ${}^{133}$Cs is included in the simulations and the optimization in the finesse is due to the extra levels in Cs-atoms, which introduces additional couplings. In general, high (low) finesse is optimal for short (long) readout times. A small cavity detuning was also included in the optimization in order to compensate for the shifts caused by the couplings to the extra levels~\cite{SM}. For a finesse of $\sim$50 and a readout time of $t\approx 200$ $\mu$s, a readout efficiency of $\eta_{\text{read}}\approx90\%$ is obtained.

\section{Errors}
Above we have focussed on the efficiency of the protocol. We will now consider the errors, which limit the performance of the system as a single photon source with memory. We find that the dominant errors are multiple excitations during the write process and the possibility of reading out atoms, which have been incoherently moved to state $\ket{1}$ by either inefficient optical pumping or wall collisions.

Multiple excitations in the write process would also create multiple quantum photons, which could in principle be discriminated from the situation with a single quantum photon if perfect single photon detection is possible. In a realistic setup there will, however, always be some finite detection probability $\eta_{\text{d}}$ and the probability of creating two excitations would introduce an error of $\sim2(1-\eta_{\text{d}})p_{\text{e}}$ to lowest order where $p_{\text{e}}\propto\int_{0}^{t_{\text{int}}}\avg{\abs{\theta_{j}(t)}^{2}}$ is the excitation probability. This error can be made arbitrarily small by simply decreasing $p_{\text{e}}$, i.e. decreasing the strength of the classical drive. This will, however, also decrease the rate of the operation, which scales as $1/p_{\text{e}}$.

\begin{figure}
\centering
\includegraphics[width=0.45\textwidth]{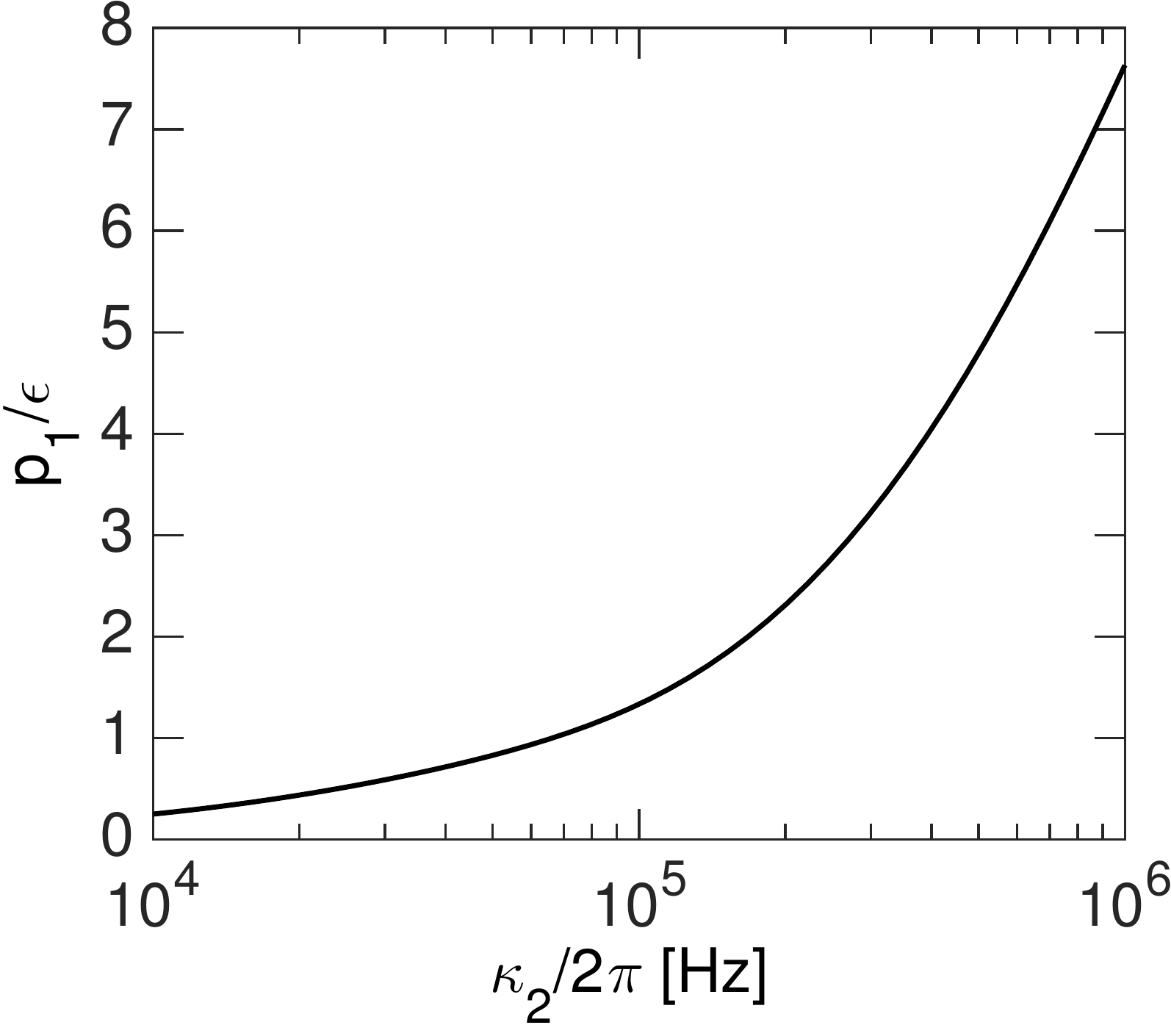}
\caption{The probability to read out incoherent photons ($p_{1}$) normalized by the fraction of atoms ($\epsilon$) that have been incoherently transferred to the readout state ($\ket{1}$). $p_{1}/\epsilon$ essentially only depends on $\tau_{read}\Gamma_{read}$ and the linewidth of the filter-cavity ($\kappa_{2}$) for the parameters that we are considering, which are $\epsilon \ll 1$, an optical depth of 168, and a finesse of the cell-cavity in the range 20-100. Furthermore, we have assumed that $\tau_{read}\Gamma_{read}=3$ which ensures a temporal filtering of the incoherent photons while keeping a high readout efficiency of the coherent photons. The plot was obtained by numerically simulating the Cs-cells used in the proof-of-principle experiment including the full level structure of the Cs-atoms.}
\label{fig:figureerror}
\end{figure}

Atoms can also be in the readout state $\ket{1}$ either by inefficient optical pumping or by wall collisions. These atoms will mainly produce \emph{incoherent} photons different in both frequency and temporal shape from the \emph{coherent} single photons originating from the symmetric excitation. The incoherent photons will have a much broader temporal and frequency profile than the coherent photons. We can thus to some extent filter them from the coherent photons by sending the light through a filter-cavity, which makes a spectral filtering, as well as having a not too long readout time $\tau_{\text{read}}$, which makes a temporal filter. In addition to the incoherent photons, atoms incoherently prepared in the wrong state can also produce coherent photons because the incoherent atoms have an overlap with the symmetric modes.  If a fraction $\epsilon$ of the atoms are transferred to the state $\ket{1}$, the probability to read out a coherent single photon from these atoms is $\epsilon\eta_{\text{read}}$. The probability $p_{1}$  to read out an incoherent photon can be found to lowest order  by assuming that an excitation is stored in any asymmetric mode instead of the symmetric Dicke mode in the perturbative expansion of $\hat{a}_{cell}$ described above. Doing the perturbative expansion, we then get a contribution to $a_{\text{cell}}$ from these incoherent excitations to the first order term $a_1$. From this, we can find the number of incoherent photons in the retrieval.  We have evaluated $p_{1}$ by numerical simulating the Cs-cells as for the readout~\cite{SM}. \figref{fig:figureerror} shows $p_{1}/\epsilon$ as a function of the linewidth ($\kappa_{2}$) of the filter-cavity.
We have assumed that $\tau_{\text{read}}=3/\Gamma_{\text{read}}$ as in \figref{fig:figure3b}. Note that this choice of readout time ensures a high readout efficiency while still making a temporal filtering of the incoherent photons since these have a smaller readout rate and hence predominantly arrive later. It is seen that $p_{1}\approx\epsilon$ for $\kappa_{2}\approx2\pi\cdot80$ kHz. With a linewidth of the filter cavity more narrow than this, the error will thus be dominated by the coherent photons which are emitted with a probability $\epsilon \eta$. Imposing this linewidth of the filter cavity for the numerical example for the readout efficiency given at the end of Sec. \ref{sec4} with a readout time of $t=200$ $\mu$s would make it drop from $\approx90\%$ to $\approx86\%$. Hence we loose only a little on the readout efficiency by filtering out the incoherent photons. Experimentally, it will be simpler to use the same filter cavity for the retrieval as for the write process, and hence it may be desirable to use a more narrow filter cavity to have an efficient write process (see Fig.~\ref{fig:figure3b}). In this case one can use a longer read out time $\tau_{\text{read}}$ to suppress loss from the filter cavity. After filtering out the incoherent photons, the remaining error is thus only due the coherent photons from atoms being incoherently prepared in the wrong state, and the error is equal to  the probability  $\epsilon$ that each atom is in  the wrong state.

\section{Conclusion and discussion}
In conclusion, we have developed a theory for motional averaging for discrete variable systems and proposed an efficient and scalable single photon source based on atomic ensembles at room temperature. We have considered a specific setup where the atomic ensemble is kept in a small cell inside a cavity and shown how both read and write efficiencies above 90\% can be achieved for a real experimental system based on Cs-atoms. The write and read processes have a timescale of $100-200$ $\mu$s, which is considerably shorter than the demonstrated quantum memory time of 10 ms~\cite{vasilakis}. To verify the essential effect described by the theory, we have performed a proof-of-principle experiment with room temperature Cs atoms contained in a microcell with spin preserving coating deposited on the walls. The measurement of the scattered light reveals long coherence time at the single photon level, resulting in a narrow peak, which is in excellent agreement with the theoretical model being used. This thus confirms the essential feature of the theory.

The room temperature cells considered here provides a promising building block for future quantum networks because of their scalability compared to cold atomic ensembles. As a particular application, we have considered a basic step of DLCZ quantum repeater with a single entanglement swap and a distance of 80 km assuming a dark count rate of 1 Hz and single photon detection efficiency of 95\% \cite{saewoo, smith}. Including various experimental imperfections such as intra-cavity losses and inefficient in/out coupling of the cavities, we estimate that a pair with fidelity $\sim80\%$ can be distributed with a rate of $\sim0.2$ Hz \cite{SM}. In this estimate, we have neglected effects from limited memory time and have assumed that a fraction of $0.5\%$ of the atoms have been incoherently transferred to the state $\ket{1}$. Note, however, that the rate of entanglement distribution can be greatly enhanced using spatially multiplexing schemes, which are possible due to the scalable nature of the room temperature cells. A particularly attractive feature of such multiplexing is that it  also decreases the necessary memory time \cite{collins}, and thus relaxes one of the most challenging requirements for long distance communication based on atomic ensembles. The microcells introduced here may thus serve as an essential building block for future photonic networks. On the other hand, for more near term applications the scalable nature of the setup will also be highly interesting for applications requiring multiple single photon inputs such as for instance photonic quantum simulators \cite{aspuru,aaronson,tillmann}.

The research leading to these results has received funding from the Lundbeck Foundation, the European Research Council under the European Union's Seventh Framework Programme (FP/2007-2013) / ERC Grant Agreement n. 306576, the ERC grant INTERFACE, the U. S. ARO under the grant W911NF-11-0235 and the EU projects MALICIA and SIQS. K. Jensen and J. Borregaard acknowledge support from the Carlsberg Foundation.

\newpage

\onecolumngrid

\section{Supplemental material: Scalable photonic network architecture based on motional averaging in room temperature gas}

In this supplemental material to the article ´``Scalable photonic network architecture based on motional averaging in room temperature gas'', we describe the details of the numerical simulations referenced in the article. We also estimate the number of classical photons, which should be filtered from the quantum photon and present the exact definition of the optical depth mentioned in the article. Finally, we estimate the errors limiting the performance of the proposed setup in a DLCZ-like quantum repeater.

\section{Numerical simulation - Write}

To justify our assumption of an exponential decay of the correlations appearing in $\avg{\abs{\theta_{j}(t)}}^{2}$ and to qualitatively characterize the write efficiency, we perform a numerical simulation of a gas of non-interacting atoms in a cell. We have based the simulation on the microcells filled with Cs-atoms, which were used in the proof-of-principle experiment. These cells have dimensions of $300$ $\mu$m $\times$ $300$ $\mu$m $\times$ $1$ cm. The cells have been placed inside a cavity with a linewidth of $\kappa_{1}\approx 2\pi \cdot 46$ MHz and both the field from the quantum photon and the classical drive are assumed to have approximately a Gaussian shape with a waist of $55$ $\mu$m. The small beam waist ensures that we can neglect cavity losses from the walls of the cell. An approximate $\Lambda$-atom can be realized in the hyperfine states of Cs with state $\ket{0}=\ket{F=4,m_{F}=4}$ and state $\ket{1}=\ket{F=3,m_{F}=3}$ in the $6^{2}S_{1/2}$ ground state manifold. The Doppler width of the atomic levels is $\Gamma_{d}\sim2\pi \cdot 225$ MHz at a temperature of $T=293$ K and we assume a detuning of $\Delta\sim4\Gamma_{d}$ from the excited level such that the effect of Doppler broadening is negligible (see below).

\begin{figure} [H]
\centering
\includegraphics[width=0.5\textwidth]{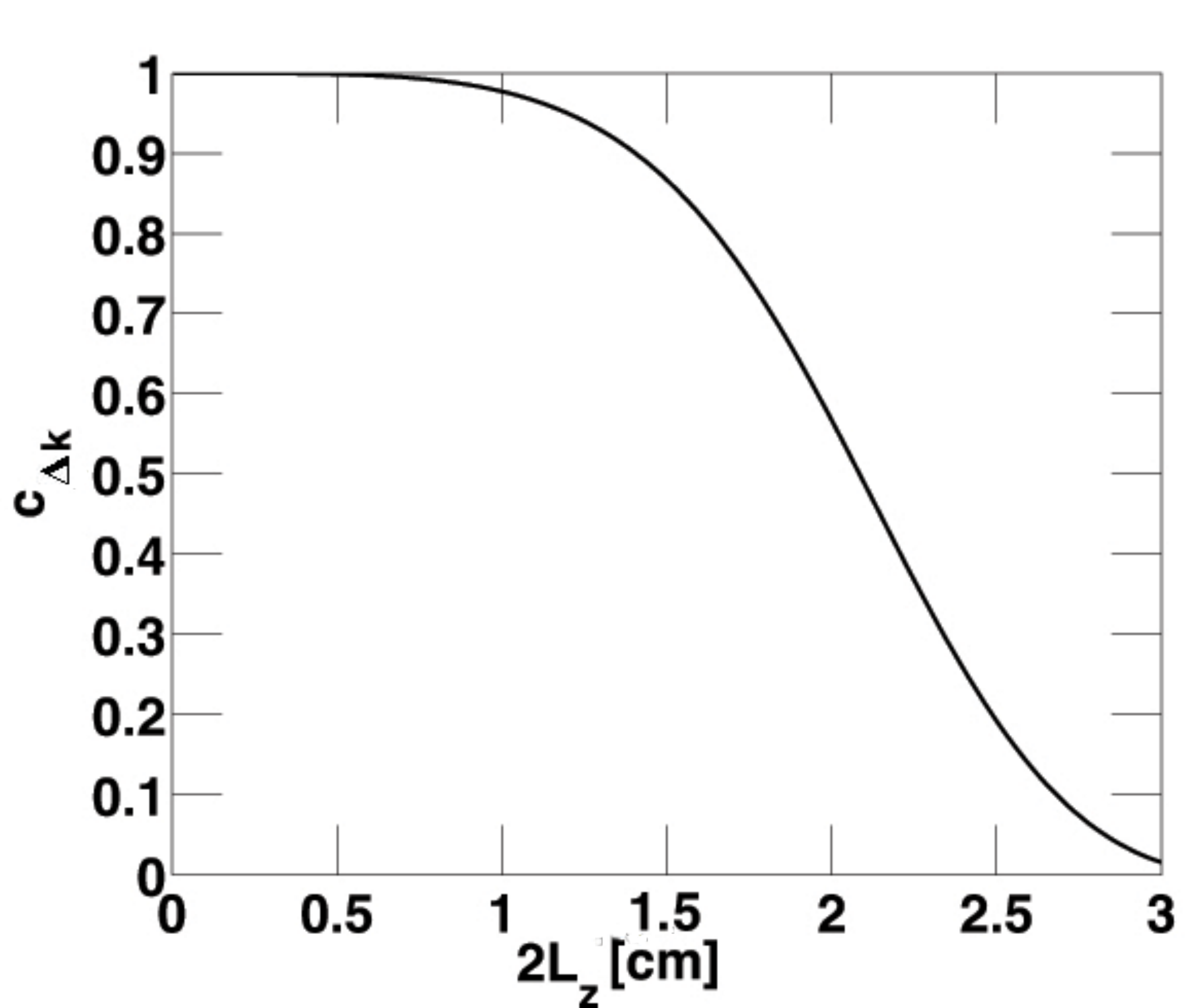}
\caption{Limitation in efficiency, $c_{\Delta k}$, from the difference in wavenumbers as a function of the cell length $2L_{z}$ for $\Delta_{k} \approx 193$ $\text{m}^{-1}$ corresponding to the $2\pi \cdot 9.2$ GHz splitting between the hyperfine ground states of ${}^{133}$Cs. The atoms are assumed to be evenly distributed in the cell. $c_{\Delta k}\gtrsim 0.97$ for $2L_{z}\lesssim 1$ cm. }
\label{fig:figureS2}
\end{figure}

Starting from Eq. (9) in the article and performing the integral over $t''$ as described in the text below, we can express $\avg{\abs{\theta_{j}(t)}}^{2}$ as
\begin{eqnarray} \label{eq:thetax21}
\avg{\abs{\theta_{j}(t)}^{2}}_{e}&=&\frac{1}{16}\int_{0}^{t}\!\!\!dt_{1}'\int_{0}^{t_{1}'}\!\!\!dt_{1}''\int_{0}^{t}\!\!\!dt_{2}'\int_{0}^{t_{2}'}\!\!\!dt_{2}''e^{-\kappa_{2}/2(t-t_{1}')}e^{-\kappa_{1}/2(t_{1}'-t_{1}'')}e^{-\kappa_{2}/2(t-t_{2}')}e^{-\kappa_{1}/2(t_{2}'-t_{2}'')} \qquad \nonumber \\
&&\times\avg{XY_{j}^{*}(t_{1}'')XY_{j}(t_{2}'')Z_{j}^{*}(t_{1}'')Z_{j}(t_{2}'')}_{e},
\end{eqnarray}
where we have defined
\begin{eqnarray}
Z_{j}(t)&=&\frac{e^{i(\Delta_{k})z_{j}(t)}-e^{-i(k_{c}+k_{q})z_{j}(t)}}{-\gamma/2+i(\Delta+k_{c}v_{z}^{(j)}(t))}+\frac{e^{-i(\Delta_{k})z_{j}(t)}-e^{i(k_{c}+k_{q})z_{j}(t)}}{-\gamma/2+i(\Delta-k_{c}v_{z}^{(j)}(t))}, \label{eq:Zeq1}\\
XY_{j}(t)&=&\Omega g e^{\frac{-2x_{j}^{2}(t)-2y_{j}^{2}(t)}{w^{2}}}.
\end{eqnarray}
Note that we have not made the assumption of $k_{c}\approx k_{q} \approx k$, as in our analytical calculations, since we have the $2\pi \cdot 9.2$ GHz splitting between the ground states, which corresponds to $k_{q}-k_{c}=\Delta _{k}\approx193$ $\text{m}^{-1}$. With a cell length of $2L_{z}=1$ cm the assumption $2\Delta_{k} L_{z} \ll 1$ is close to being violated and we shall therefore not make this assumption. The extra terms $ \propto e^{\pm i\Delta_{k} z_{j}}$ will approximately result in a factor of
\begin{equation}
c_{\Delta k}=\frac{\avg{\cos\left(\Delta_{k} z_{j}\right)}^{2}}{\avg{\cos\left(\Delta_{k} z_{j}\right)^{2}}},
\end{equation}
which should be multiplied with the analytical expression for the write efficiency, which was obtained assuming $\Delta_{k}=0$. \figref{fig:figureS2} shows how $c_{\Delta k}$ depends on the length of the cell assuming that the atoms are equally distributed in the entire cell. It is seen from \figref{fig:figureS2} that as long as the length of the cell is $2L_{z}\lesssim1$ cm then $c_{\Delta k} \gtrsim 0.97$ for $\Delta_{k}\approx193$ $\text{m}^{-1}$ and hence the frequency difference between the quantum and classical fields does not significantly degrade the write efficiency. In all our numerical simulations, we, however, keep the terms $\propto e^{\pm i\Delta_{k} z_{j}}$ for completeness.

The correlations appearing in $\avg{XY_{j}^{*}(t_{1}'')XY_{j}(t_{2}'')Z_{j}^{*}(t_{1}'')Z_{j}(t_{2}'')}_{e}$ (see Eq.~\eqref{eq:thetax21}) depend on the time difference $\abs{t''_{1}-t''_{2}}$ and we therefore introduce the shorthand notation $\avg{XY_{j}^{*}(t_{1}'')XY_{j}(t_{2}'')Z_{j}^{*}(t_{1}'')Z_{j}(t_{2}'')}_{e}=\avg{XY,Z}_{e}(t''_{1}-t''_{2})$. We change to the variables  $u=t''_{1}+t''_{2}$ and $s=t''_{1}-t''_{2}$ and by changing the order of integration, we can perform the integrals over $t, t'_{1},t'_{2}$ and $u$. To obtain the write efficiency $\eta_{\text{write}}$, we need to perform an additional integration over $t$ (see Eq.~(10) in the article). We are therefore left with
\begin{equation} \label{eq:supthetaeff}
\int_{0}^{t_{\text{int}}}\avg{\abs{\theta(t)}^{2}}_{e}dt=\int_{0}^{t_{\text{int}}}h(t_{\text{int}},\kappa_{1},\kappa_{2},s) \avg{XY,Z}_{e}(s)ds,
\end{equation}
where $h(t_{\text{int}},\kappa_{1},\kappa_{2},s)$ is a function of $s$ obtained by performing the integrals over $t, t'_{1},t'_{2}$ and $u$. We can evaluate the integral over $s$ numerically by simulating the correlations $\avg{XY,Z}_{e}(s)$. Since the atoms do not interact with each other, we independently simulate the motion of $N=5000$ atoms through the cell and evaluate the correlations of atoms at points separated in time by $s$. Finally, we average over many realizations. At room temperature we assume that the motion of the atoms is classical. Furthermore, we assume that the atoms are evenly distributed in the cell and that their velocity distribution is described by the Maxwell Boltzmann distribution at a temperature $T=293$ K. We assume that the atoms are re-thermalized completely after every collision with the walls of the cell but qualitatively similar results are obtained for a ballistic model without thermalization. For the ballistic model, the $Z_{j}$ parts of the couplings in principle do not average down. This could lead to effects not averaged away by using narrow filter cavities. However, we are far detuned compared to the Doppler width of the atoms and the cavity fields are standing waves, which can be viewed as the superposition of two counter propagating waves. As a result, the effect of the velocity fluctuations of the atoms cancel and the fluctuations in the $Z_{j}$ terms are greatly suppressed. This is in contrast to what happens in ensemble based schemes with a laser coming from one side where Doppler effects do not go away by working far off resonance \cite{gorshkov}. Consequently, we expect similar results for the ballistic model as for the model with complete thermalization. We have explicitly verified this by repeating our simulations with the ballistic model, which lead to more or less identical results as for the model with thermalization (the two simulations cannot be distinguished due to the noise from the random nature of the simulations). The result of a simulation with thermalization is seen in \figref{fig:figuretS1a}, which shows how the correlations decay as a function of $s$ such that for $s\to\infty$, we have $\avg{XY,Z}_{e}(s) \to \abs{\avg{XY}_{e}}^{2}\abs{\avg{Z}_{e}}^{2}$. This enables us to introduce a maximal cutoff, $s_{\text{max}}$, in the numerical integral appearing in Eq.~\eqref{eq:supthetaeff}, above which, the correlations have effectively vanished. As a result, we can semianalytically evaluate $\eta_{\text{write}}$ for an arbitrary pulse length $t_{\text{int}}$ without additional numerical difficulty. Note that \figref{fig:figuretS1a} also shows that the exponential model of the decay of the correlations assumed in our analytical calculations is a good approximation.

Based on our simulations of $\avg{XY,Z}_{e}(s)$, we have also estimated the power spectral density, PSD in the proof of principle experiment. The PSD was measured (see Fig. 2 in the article and \figref{fig:supx} below) by measuring the Faraday rotation of light transmitted through the cell. In these measurements the light is emitted by the scattering between different $m_F$ states in an applied magnetic field. As a result, the signal is modulated at the Larmor precession frequency. We therefore consider the modulated power spectral density
\begin{eqnarray}
\text{PSD}(f)&\propto&\frac{1}{t_{\text{int}}^{2}}\int_{0}^{t_{\text{int}}}dt\int_{0}^{t_{\text{int}}}dt'\avg{XY,Z}_{e}(t-t')e^{2i\pi f (t-t')} \\
&\propto&\frac{1}{t_{\text{int}}^{2}}\int_{0}^{t_{\text{int}}}dt\int_{0}^{t_{\text{int}}}dt'\abs{\avg{XY}_{e}}^{2}\abs{\avg{Z}_{e}}^{2}e^{2i\pi f (t-t')}  \nonumber \\
&&+\frac{1}{t_{\text{int}}^{2}}\int_{0}^{t_{\text{int}}}dt\int_{0}^{t_{\text{int}}}dt'(\avg{XY,Z}_{e}(t-t')-\abs{\avg{XY}_{e}}^{2}\abs{\avg{Z}_{e}}^{2})e^{2i\pi f (t-t')}\\
&\propto&\delta_{f,0}+\frac{1}{t_{\text{int}}^{2}}\int_{0}^{t_{\text{int}}}dt\int_{0}^{t_{\text{int}}}dt'(\avg{XY,Z}_{e}(t-t')-\abs{\avg{XY}_{e}}^{2}\abs{\avg{Z}_{e}}^{2})e^{2i\pi f (t-t')}
\end{eqnarray}
where $f$ is the frequency and $\delta_{f,0}$ is the Kronecker delta function. Note that we have assumed that  the coherent light-atom interaction gives a contribution of $\frac{1}{t_{\text{int}}^{2}}\int_{0}^{t_{\text{int}}}dt\int_{0}^{t_{\text{int}}}dt'\abs{\avg{XY}_{e}}^{2}\abs{\avg{Z}_{e}}^{2}e^{2i\pi f (t-t')}=\delta(f)$ to the PSD. In the proof-of-principle experiment, the atoms are subject to a magnetic field, which makes the atomic spins precess around the mean spin direction with a Larmor frequency of 823.8 kHz. The Raman scattered part of the PSD is thus centered around this frequency. However, the measured PSD also contains both the shot noise of the light and electronic noise from the measurement equipment. Since we are only interested in the signal from the atomic interaction, this noise is isolated by performing a second measurement at a higher Larmor frequency (2594 kHz) and subtract it from the first measurement. The higher Larmor frequency is chosen such that the two atomic signals are well separated in frequency. In the simulated PSD, we have fitted a Lorentzian to the broad feature centered at 823.8 kHz and shifted it to be centered at 2594 kHz. This has then been subtracted from the simulated data to include the substraction of the two signals in the experiment.

To validate that the proof-of-principle experiment is probing the theory in the right limit of single photon Raman scattering, we estimate the number of Raman photons scattered over the relevant pulse length. This number can be found from the ratio of the height of the central peak in the PSD to the shot noise of light.
In the experiment, the balanced polarimetry measures the amplitude quadrature of the field in the polarisation axis perpendicular to the drive field $\hat{a}_{\text{sc}}$. Neglecting proportionality factors irrelevant for this calculation, the PSD at a discrete frequency $f$ is given by: $\text{PSD}(f)   \propto \tilde{\mathcal{X}}^{\dag}(f) \tilde{\mathcal{X}}(f)$, where:
\begin{eqnarray}
\tilde{\mathcal{X}}(f) & \propto & \int_0^{t_{\text{ms}}}  \left[ \hat{a}_{\text{sc}}(t) + \hat{a}^{\dag}_{\text{sc}}(t) \right] e^{2i\pi f t} dt = \hat{a}_{\text{sc}}(f)+\left[ \hat{a}_{\text{sc}}(-f) \right]^\dag.
\end{eqnarray}
Here $\hat{a}_{\text{sc}}(f) = \int_0^{t_{\text{ms}}} \hat{a}_{\text{sc}}(t) e^{2 i \pi f} dt$ and $t_{\text{ms}}$ is the measurement time. It can then be shown that the PSD takes the form:
\begin{equation}
\text{PSD}(f)  \propto 1+\left[ \hat{a}_{\text{sc}}(-f) \right]^\dag \left[ \hat{a}_{\text{sc}}(-f) \right]+\left[ \hat{a}_{\text{sc}}(f) \right]^\dag \left[ \hat{a}_{\text{sc}}(f) \right]. \label{eq:supPSDNumPhotons}
\end{equation}
The unity term in the left-hand part of the above relationship is the photon shot noise contribution, present even when there is no scattered field; the second and third terms represent the Raman-scattered number of photons in the lower (Stokes line) and upper (anti-Stokes line) sideband respectively. In \figref{fig:supx} the PSD recorded with a pulse duration of $t_{\text{ms}}\approx 16$~$\mu$s is plotted. The peak height at the Larmor frequency due to the  Raman scattering has equal contributions from the Stokes and the anti-Stokes lines. The number of coherently scattered photons during the measurement pulse is the ratio of the peak height excluding the photon shot noise and the incoherent photons (denoted with a dotted lines in \figref{fig:supx}) to the photon shot noise level, which corresponds to a power of one photon per unit bandwidth as shown in Eq.~\eqref{eq:supPSDNumPhotons}. From \figref{fig:supx} we find that for the proof-of-principle experiment approximately $1.2$ photons are coherently scattered in the Stokes line during the 16~$\mu$s pulse. This corresponds to approximately $8$ photons over a pulse duration of 106~$\mu$s that can lead to an efficient write step according to our proposal. Although this is slightly higher than the single photon required for the protocol, we expect the spectrum to be independent of intensity at these light levels.

In the above calculation it is important to estimate the photon shot noise level. For this, the electronic noise, as measured with no light at the detector, is removed from the recorded PSD and the spectrum is corrected for the detector frequency response. The photon shot noise then corresponds to the level where the PSD levels off at frequencies a few linewidths of the broad Lorentzian away from the Larmor frequency; at these frequencies the power from the Raman scattered photons is negligible and the PSD is determined only by the photon shot noise. The photon shot noise is also verified by performing balanced polarimeter detection of light that has not interacted with the atomic ensemble and with power equal to the one as the drive light in the experiment. We note that the detector response was characterized by such a measurement.

\begin{figure}
\centering
\includegraphics[width=0.5\textwidth]{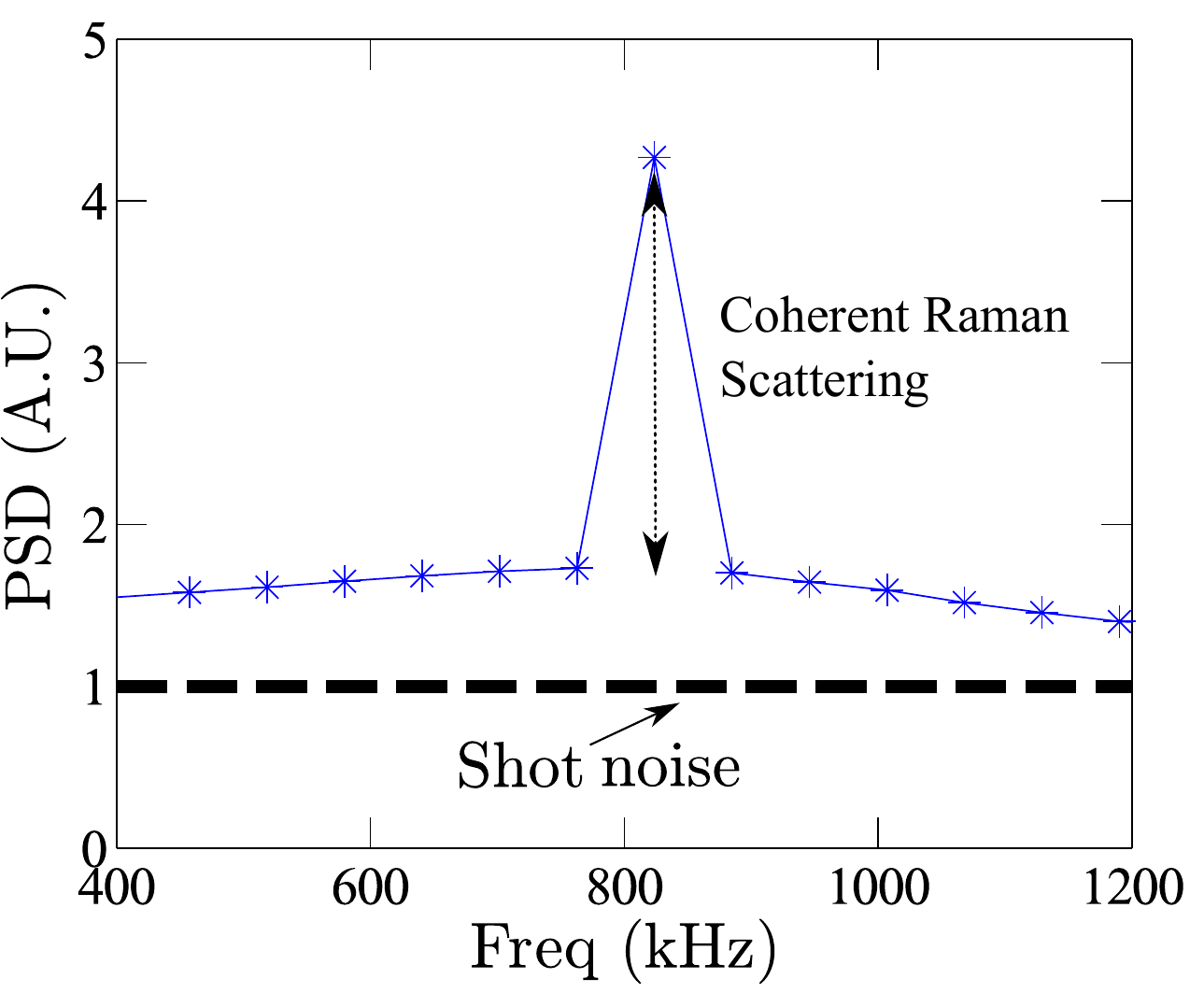}
\caption{Experimental data of the power spectral density (PSD) of the Raman scattered light measured in the proof-of-principle experiment. These data were obtained for a pulse length of $\sim16$~$\mu$s. The electronic noise has been removed from the data and the shot noise has been normalized to unity. The number of coherent Raman-scattered photons, equally distributed in the Stokes and anti-Stokes lines, can be estimated by the comparing the height of the central peak (dotted line) to the shot noise of light. For the proof of principle experiment, the number of coherently scattered photons in the antiStokes sideband over the 16~$\mu$s of the pulse is estimated to be $\sim 1.2 $~photons.}
\label{fig:supx}
\end{figure}

\section{Number of photons}

As mentioned in the article, the purpose of the filter cavity is both to increase the averaging time and to filter the quantum photon from the classical photons. We will now estimate the number of classical photons, which needs to be filtered from the single quantum photon. In order to do this, we need to characterize the ensemble, which we do by introducing the optical depth $d$.

\begin{figure}
\centering
\subfloat {\label{fig:figuretS1a}\includegraphics[width=0.5\textwidth]{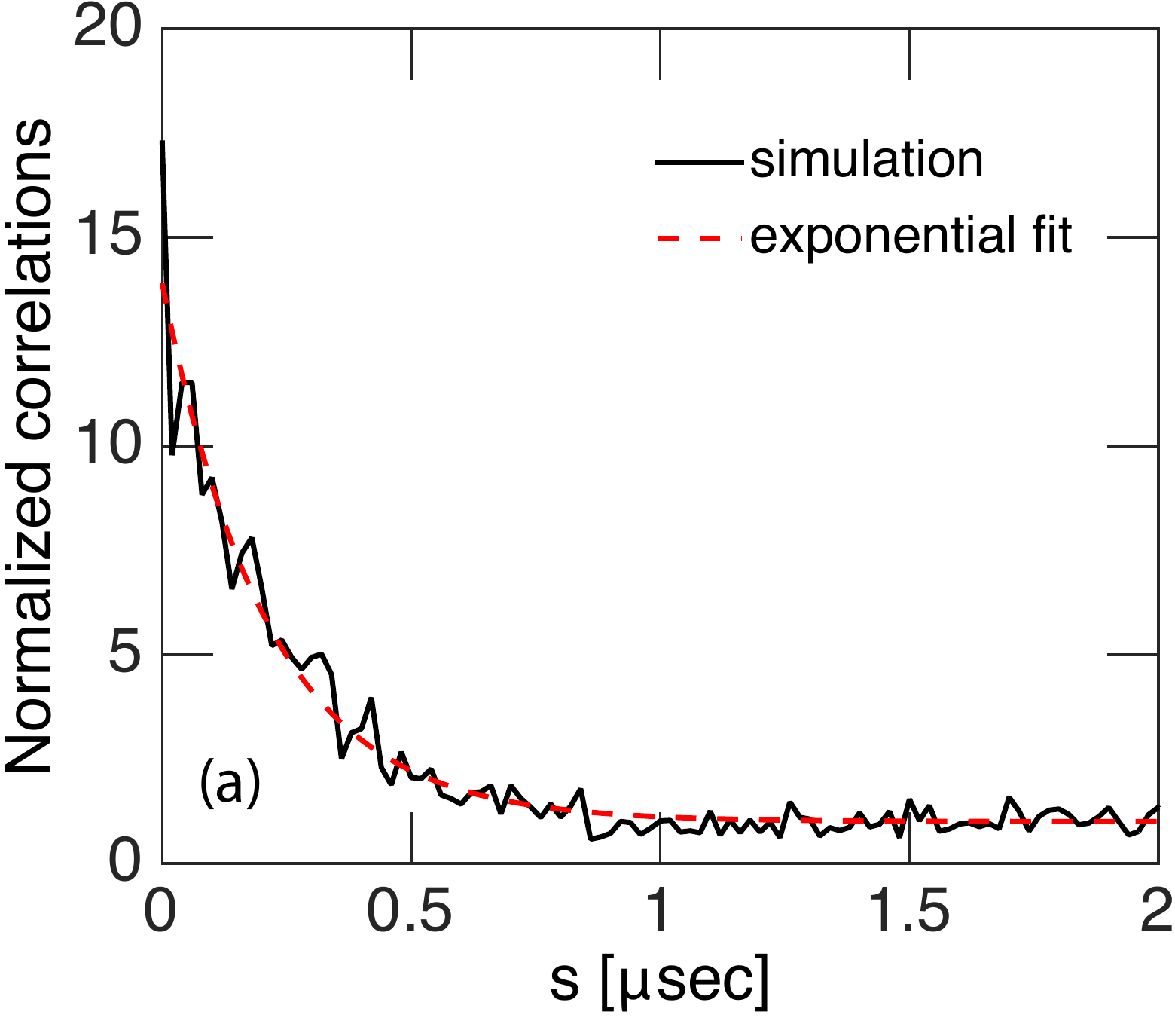}}
\subfloat{\label{fig:figuretS1b}\includegraphics[width=0.44\textwidth,height=3.2in]{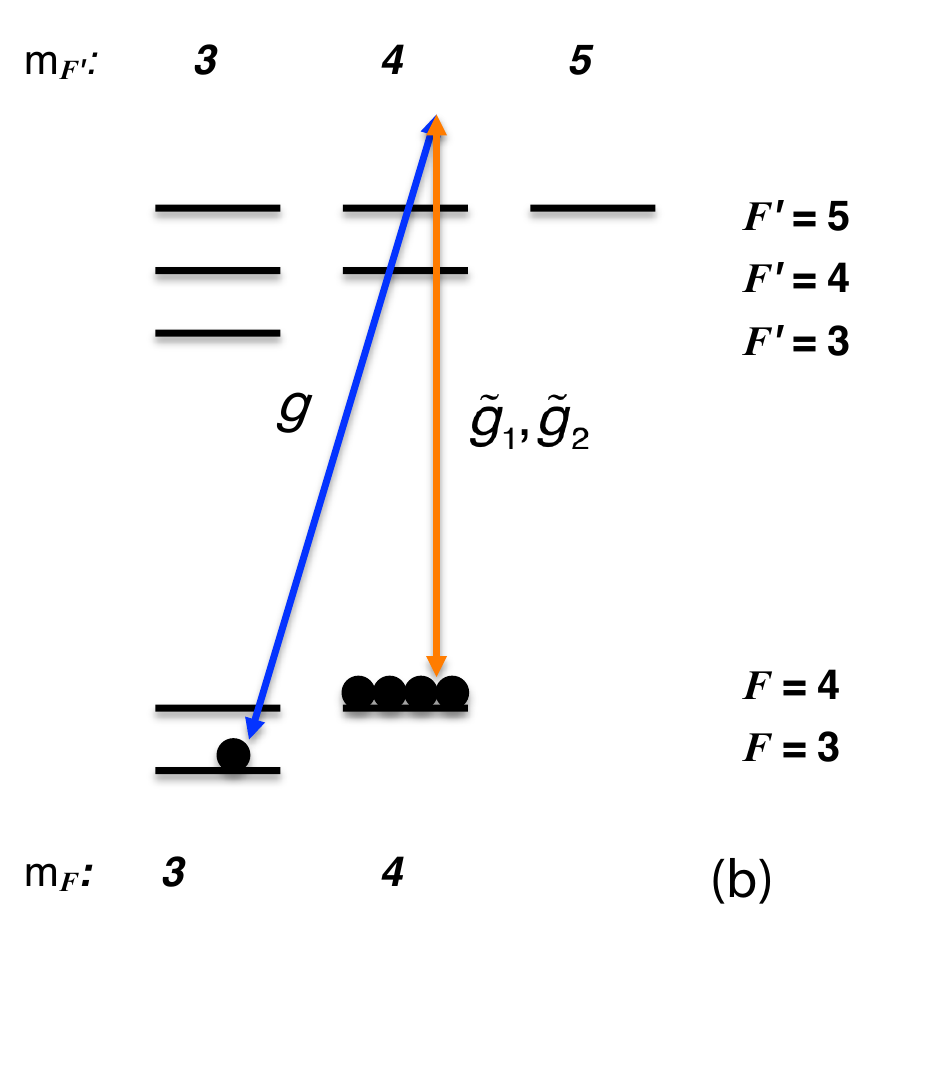}}
\caption[Simulation of atomic correlations]{(a) Simulation of the correlations $\avg{XY,Z}_{e}(s)$. The correlations are normalized to be unity for $s\to\infty$ where there are no correlations and $\avg{XY,Z}_{e}(s)\to\abs{\avg{XY}_{e}}^{2}\abs{\avg{Z}_{e}}^{2}$. The data from the simulation have been fitted with an exponential model validating our assumption of an exponential decay of the correlations. The fit gives a decay rate of $\Gamma=2\pi\cdot0.75$ MHz corresponding to $\Gamma\sim1.3 v_{\text{thermal}}/w$. (b) Sketch of the $6^{2}S_{1/2}$ and $6^{2}P_{3/2}$ hyperfine levels in ${}^{133}$Cs. A $\Lambda$-atom is realized with $\ket{0}=\ket{F=4,m_{F}=4}$, $\ket{1}=\ket{F=3,m_{F}=3}$ as ground states in $6^{2}S_{1/2}$ and $\ket{e_{1}}=\ket{F'=4,m_{F'=4}}$ as the excited level in $6^{2}P_{3/2}$. To characterize the optical depth, we consider the transitions $\ket{0}\to\ket{e_{1}}$ and $\ket{0}\to\ket{e_{2}}=\ket{F'=5,m_{F'}=4}$ characterized by $\tilde{g}_{1}$ and $\tilde{g}_{2}$, respectively.  }
\label{fig:figuretS12}
\end{figure}

To obtain an expression for the optical depth, we assume that we are working with the previously mentioned Cs-cells. The relevant level structure is shown in \figref{fig:figuretS1b}.
In the write process, the classical drive is applied on the transitions between the ground state $\ket{0}=\ket{F=4,m_{F}=4}$ and the excited states $\ket{e_{1}}=\ket{F'=4,m_{F'}=4}$ and $\ket{e_{2}}=\ket{F'=5,m_{F'}=4}$ characterized by $\tilde{g}_{1}$ and $\tilde{g}_{2}$, respectively.  The quantum photon is created on the transition $\ket{e_{1}}\to\ket{1}=\ket{F=3,m_{F}=3}$ characterized by $g$. Note that with this field configuration, the cell-cavity in principle also mediate the transition $\ket{F=4,m_{F}=4} \to \ket{F=4,m_{F}=3}$ in the write setup but this transition is suppressed by the $2\pi\cdot9.2$ GHz splitting between the ground states, which makes the corresponding photon non-resonant with the subsequent filter-cavity. This transition will, therefore, never give a click in the detector. Since the interaction is only a perturbation to the system, we can therefore neglect this transition in our numerical simulations.  All atoms in the ensemble are initially pumped to the ground state $\ket{0}$ and, in order to characterize the optical depth, we assume that the transition characterized by $g$ is non-driven and ignore any cavity (Purcell) enhancement of the corresponding decay. The cavity field thus only couples $\ket{0}\to\ket{e_{1}}$ and $\ket{0}\to\ket{e_{2}}$ with coupling constants $\tilde{g}_{1}$ and $\tilde{g}_{2}$, respectively (see \figref{fig:figuretS1b}).

The equations of motion for the cavity field, $\hat{a}_{\text{cav}}$, and the relevant atomic operators in a suitable rotating frame are
\begin{eqnarray}
\dot{\hat{a}}_{\text{cav}}&=&-(\kappa/2)\hat{a}_{\text{cav}}+i\sum_{j=1}^{N}\left[\tilde{g}_{1}^{(j)}(t)\hat{\sigma}_{e_{1}0}^{(j)}+\tilde{g}_{2}^{(j)}(t)\hat{\sigma}_{e_{2}0}^{j}\right] \label{eq:eomd1} \\
\dot{\hat{\sigma}}_{e_{1}0}^{(j)}&=&-(\gamma_{1}/2-i\Delta_{1})\hat{\sigma}_{e_{1}0}^{(j)}+i\tilde{g}_{1}^{(j)}(t)\hat{a}_{\text{cav}} \label{eq:eomd2}\\
\dot{\hat{\sigma}}_{e_{2}0}^{(j)}&=&-(\gamma_{2}/2-i\Delta_{2})\hat{\sigma}_{e_{2}0}^{(j)}+i\tilde{g}_{2}^{(j)}(t)\hat{a}_{\text{cav}} \label{eq:eomd3},
\end{eqnarray}
where $\sigma_{e_{l}0}^{(j)}=\ket{e_{l}}_{j}\bra{0}$ ($l=1,2$) and we have assumed that $\sigma_{e_{l}e_{l}}^{(j)}-\sigma_{00}^{(j)}\approx-1$. For simplicity, we have assumed the couplings ($\tilde{g}$) to be real. $\Delta_{1}=\omega_{1}-\omega_{\text{cav}}$ ($\Delta_{2}=\omega_{2}-\omega_{\text{cav}}$) is the detuning of $\ket{e_{1}}$ ($\ket{e_{2}}$), while $\gamma_{1}$ ($\gamma_{2}$) is the corresponding decay rate. Here $\omega_{1}$ ($\omega_{2}$) is the frequency associated with the atomic level and $\omega_{\text{cav}}$ is the frequency of the cavity field. Formally integrating Eqs. \eqref{eq:eomd2}-\eqref{eq:eomd3}, assuming that $\sigma_{e_{1}0}^{(j)}=\sigma_{e_{2}0}^{(j)}=0$ at time $t=0$, and inserting the resulting expression for $\sigma_{e_{1}0}^{(j)}$ and $\sigma_{e_{2}0}^{(j)}$ into Eq.~\eqref{eq:eomd1} gives
\begin{eqnarray}\label{eq:field1}
\dot{\hat{a}}_{\text{cav}}&=&-(\kappa/2)\hat{a}_{\text{cav}}-\sum_{j=1}^{N}\Bigg[\tilde{g}_{1}^{(j)}(t)\int_{0}^{t}e^{-(\gamma_{1}-i\Delta_{1})(t-t')}\tilde{g}_{1}^{(j)}(t)\hat{a}_{\text{cav}}(t')\text{d}t'\nonumber \\
&&+\tilde{g}_{2}^{(j)}(t')\int_{0}^{t}e^{-(\gamma_{2}-i\Delta_{2})(t-t')}\tilde{g}_{2}^{(j)}(t')\hat{a}_{\text{cav}}(t')\text{d}t'\Bigg],
\end{eqnarray}
where we have explicitly written the time dependence of $\hat{a}_{\text{cav}}$ inside the integrals. We evaluate the integrals in Eq.~\eqref{eq:field1} assuming that we can treat $\hat{a}_{\text{cav}}(t')$ as a constant in time and move it outside the integrals. Furthermore, we assume that $\tilde{g}_{l}^{(j)}(t')=\tilde{g}_{l,xy}^{(j)}(t)\sin(k(z_{j}(0)+v_{z}^{(j)}(0)t'))$ similar to the procedure described in the article. Note that $k$ is the wavenumber associated with the cavity field while $z_{j}(0)$ ($v_{z}^{(j)}$) is the $z$-part of the position (velocity) of the $j$'th atom. After evaluating the integrals, we obtain
\begin{equation} \label{eq:field2}
\dot{\hat{a}}_{\text{cav}}=-(\kappa/2)\hat{a}_{\text{cav}}+\frac{\hat{a}_{\text{cav}}}{4}\sum_{j=1}^{N}\left[\abs{\tilde{g}_{1,xy}^{(j)}(t)}^{2}Z_{j}(\Delta_{1},\gamma_{1},k)+\abs{\tilde{g}_{2,xy}^{(j)}(t)}^{2}Z_{j}(\Delta_{2},\gamma_{2},k)\right],
\end{equation}
where we have adiabatically eliminated the optical coherence and have rewritten $Z_{j}(t)$ defined in Eq.~\eqref{eq:Zeq1} to
\begin{equation}
Z_{j}(\Delta,\gamma,k)=\frac{e^{2ikz_{j}(t)}-1}{\gamma/2+i(kv_{j}(0)-\Delta)}-\frac{1-e^{-2ikz_{j}(t)}}{\gamma/2-i(kv_{j}(0)+\Delta)},
\end{equation}
such that $\gamma, \Delta$, and $k$ become variable parameters. We now perform an ensemble average of Eq.~\eqref{eq:field2} assuming that the atoms are evenly distributed in the cell and that their velocity distribution follows a Maxwell Boltzmann distribution, as previously considered. Furthermore, we assume that the $xy$-dependence of the couplings are Gaussians similar to Eqs.~(11)-(12) in the article and that we are detuned far from the Doppler width of the atoms. This results in
\begin{equation}
\dot{\hat{a}}_{\text{cav}}=-(\kappa/2)\hat{a}_{\text{cell}}-\frac{\hat{a}_{\text{cell}}N}{4}\left[\frac{\abs{\tilde{g}_{1}}^{2}\gamma_{1}}{\gamma^{2}_{1}/4+\Delta^{2}_{1}}+\frac{\abs{\tilde{g}_{2}}^{2}\gamma_{2}}{\gamma^{2}_{2}/4+\Delta^{2}_{2}}\right]\frac{\pi}{8}\frac{w^{2}}{L^{2}}+i[\ldots] \label{eq:field3t},
\end{equation}
where the imaginary part is contained in $[\ldots]$. The second term in Eq.~\eqref{eq:field3t} is identified as the single pass optical depth, $\tilde{d}$, divided by the cavity round trip time, $\tau$, where $\exp(-\tilde{d})$ is the attenuation of the light field after passing through the ensemble. Since $\tilde{d}$ depends on, e.g., the detuning, it is not a direct characterisation of the ensemble. Instead, in analogy with Ref.~\cite{gorshkov3}, we characterise the ensemble by $d$, the hypothetical optical depth, which would be obtained for resonant fields in the absence of Doppler broadening and hyperfine interaction, i.e., Eq.~\eqref{eq:field3t} with $\Delta_1=\Delta_2=0$. Furthermore, we assume that $\gamma_{1}=\gamma_{2}=\gamma$ such that the optical depth is
\begin{equation} \label{eq:opticaldepth}
d=\frac{N\tau}{\gamma}\left(\abs{\tilde{g}_{1}}^{2}+\abs{\tilde{g}_{2}}^{2}\right)\alpha_{xy}
\end{equation}
where we have defined the factor $\alpha_{xy}=\frac{\pi}{8}\frac{w^{2}}{L^{2}}$. Note that Eq.~\eqref{eq:opticaldepth} can be rewritten to the following well known formula for the optical depth \cite{jackson}
\begin{equation} \label{eq:opticaldepth2}
d=6\pi\frac{N}{(2L)^{2}}\tilde{\lambda}^{2}\left(\frac{\gamma_{1,0}+\gamma_{2,0}}{\gamma}\right),
\end{equation}
where $2L$ is the transverse size of the cell, $\tilde{\lambda}=\lambda/2\pi$ is the rescaled wavelength of the light, and $\gamma_{s,0}$ is the spontaneous decay rate of level $e_{s}$ back to $\ket{0}$ (s=1,2). The optical depth can also be related to the Faraday rotation angle, $\theta_{F}$, which is typically measured in experiments and used to estimate the number of atoms, $N$ in the ensemble \cite{hammerer,kasper}. For ${}^{133}$Cs, the relation between $\theta_{F}$ and $N$ is \cite{kasper}
\begin{equation}  \label{eq:faraday}
N=\abs{\frac{32\pi L^{2} \theta_{F} \Delta_{2}}{a_{1}(\Delta_{2}) \gamma \lambda^{2}}}
\end{equation}
where $a_{1}(\Delta_{2})$ is the vector polarisability given by
\begin{equation}
a_{1}(\Delta_{2})=\frac{1}{120}\left(-\frac{35}{1-\Delta_{3'5'}/\Delta_{2}}-\frac{21}{1-\Delta_{4'5'}/\Delta_{2}}+176\right),
\end{equation}
with $\Delta_{x'5'}$ denoting the hyperfine splitting between level $F'=x$ and $F'=5$. Combining Eq.~\eqref{eq:opticaldepth2} and Eq.~\eqref{eq:faraday} gives the following relation between $d$ and $\theta_{F}$
\begin{equation}
d=\abs{\frac{12\Delta_{2} \theta_{F}}{a_{1}(\Delta_{2})}\frac{\gamma_{1,0}+\gamma_{2,0}}{\gamma^{2}}}
\end{equation}
For the cells used in the proof-of-principle experiment, the Faraday rotation angle has been measured to be $4.4^{\text{o}}$ for a detuning of $\Delta_{2}=2\pi\cdot 850$ MHz. This translates into an optical depth of $d\approx168$.

Having defined the optical depth, we can now estimate the number of classical photons that need to be filtered from the quantum photon. The field at the detector (see Fig. 1b in the article) is described by the operator $\hat{a}$ in Eq.~(8) in the article. Assuming a length, $t_{\text{int}}$, of the write pulse, we estimate the average number of quantum photons, $N_{\text{quant}}$ at the detector as
\begin{equation} \label{eq:nquant}
N_{\text{quant}}=\avg{\hat{a}^{\dagger}\hat{a}\cdot t_{\text{int}}}=\frac{1}{16}\kappa_{2}^{2}\kappa_{1}N\avg{\abs{\theta_{j}}^{2}}_{e},
\end{equation}
where we have used that $\avg{\abs{\theta_{j}}^{2}}_{e}$ is independent of time as shown in Eq.~(14) in the article. Note that $\avg{\abs{\theta_{j}}^{2}}_{e}\propto\abs{g}^{2}\abs{\Omega}^{2}$ and we estimate the number of classical photons contained in the write pulse as $N_{\text{clas}}\sim\abs{\Omega}^{2}t_{\text{int}}\kappa_{1}/(4\abs{\tilde{g}}^{2})=\abs{\Omega}^{2}t_{\text{int}}\kappa_{1}/(4\beta\abs{g}^{2})$, where $\beta=\abs{\mu_{\tilde{g}}}^{2}/\abs{\mu_{g}}^{2}$ is the ratio between the Clebsch-Gordan coefficients ($\mu$) of the transitions characterized by $\tilde{g}$ and $g$ (see \figref{fig:figuretS1b}). From Eq.~\eqref{eq:nquant}, we then get
\begin{equation}
N_{\text{clas}}\sim\frac{N_{\text{quant}}}{N}\frac{4}{\abs{g}^{4}\avg{\abs{\theta}^{2}}\beta\kappa_{2}^{2}}.
\end{equation}
The number of classical photons that needs to be filtered is finally estimated by setting $N_{\text{quant}}=1$. Using Eqs. \eqref{eq:opticaldepth}-\eqref{eq:opticaldepth2}, we can express $N_{\text{clas}}$ in terms of the optical depth and the finesse of the cell-cavity, defined as $\mathcal{F}=2\pi/(\tau\kappa_{1})$, where $\tau$ is the cavity roundtrip time. Furthermore, we assume that $\avg{\abs{\theta_{j}}^{2}}_{e}\approx\abs{\avg{\theta_{j}}_{e}}^{2}$ such that the number of classical photons can be estimated as
\begin{equation}
N_{\text{clas}}\sim\frac{8\pi\beta_{2}^{2}L^{2}\Delta^{2}}{3\beta\tilde{\lambda}^{2}\gamma(\gamma_{1}+\gamma_{2})}\frac{1}{d\mathcal{F}^{2}},
\end{equation}
where we have expanded the expression for $\abs{\avg{\theta_{j}}}^{2}$ in the limit of large detuning. $\beta_{2}=\frac{\abs{\mu_{g1}}^{2}+\abs{\mu_{g2}}^{2}}{\abs{\mu_{g}}^{2}}$ is the ratio between the Clebsch-Gordon coefficients of the transitions characterized by $\tilde{g}_{1},\tilde{g}_{2}$ and $g$ in \figref{fig:figuretS1b}. For the experimental Cs-cells and a detuning of $\Delta=2\pi\cdot898$ MHz, we find that $N_{\text{clas}}\sim\frac{7.4\cdot10^{11}}{d\mathcal{F}^{2}}$. With $d=168$ and $\mathcal{F}=100$ this gives $N_{\text{clas}}=4.4\cdot10^{5}$. Since the quantum and classical field differ both in polarisation and frequency, this level of filtering is expected to be easily achieved using using a combination of both polarisation filtering and the filter-cavity.

\section{readout}
In this section, we give the expressions for the second order correction to the readout efficiency and present the details of the simulations. For the second order term $\eta_{\text{read},2}$ we find
\begin{eqnarray} \label{eq:eta2tr}
\eta_{\text{read},2}&=&2\kappa_{1}\text{Real}\Bigg(\int_{0}^{\tau_{\text{read}}}\!\!\!dt\int_{0}^{t}\!\!\!dt'\int_{0}^{t'}\!\!\!dt''\frac{\sqrt{N}\bar{\mathcal{B}}^{*}}{2\abs{\mathcal{D}}}e^{\text{Real}(\bar{\mathcal{A}}+\bar{\mathcal{C}})t}\left(e^{\frac{1}{2}\sqrt{\mathcal{D}}^{*}t}-e^{-\frac{1}{2}\sqrt{\mathcal{D}}^{*}t}\right)e^{\bar{\mathcal{C}}(t'-t'')}e^{-\frac{1}{2}(\bar{\mathcal{A}}+\bar{\mathcal{C}})(t'-t'')} \nonumber \\
&&\times \Bigg(e^{\frac{1}{2}\sqrt{D}t}\!\left(\!e^{\frac{-1}{2}\sqrt{D}(t'-t'')}\!-\!e^{\frac{-1}{2}\sqrt{D}(t'+t'')}\!\right)\!\left(\!\left[\left(\bar{\mathcal{A}}\!-\!\bar{\mathcal{C}}\!+\!\sqrt{\mathcal{D}}\right)\!N\avg{\delta\mathcal{B}_{j}(t')\delta\mathcal{B}_{j}(t'')}_{e}\!+\!2\bar{\mathcal{B}}N\avg{\delta\mathcal{C}_{j}(t')\delta\mathcal{B}_{j}(t'')}_{e}\right]\!\frac{\sqrt{N}\bar{\mathcal{B}}}{\sqrt{\mathcal{D}}}\!\right) \nonumber \\
&&+e^{\frac{1}{2}\sqrt{D}(t+t''-t')}\!\Bigg(\!\Big[\left(\bar{\mathcal{A}}\!-\!\bar{\mathcal{C}}\!+\!\sqrt{\mathcal{D}}\right)\!N\avg{\delta\mathcal{B}_{j}(t')\delta\mathcal{C}_{j}(t'')}_{e}\!+\!2\bar{\mathcal{B}}\sqrt{N}\avg{\delta\mathcal{C}_{j}(t')\delta\mathcal{C}_{j}(t'')}_{e}\Big]\!\frac{\bar{\mathcal{A}}\!-\!\bar{\mathcal{C}}\!+\!\sqrt{\mathcal{D}}}{2\sqrt{\mathcal{D}}}\!\Bigg) \nonumber \\
&&+e^{\frac{1}{2}\sqrt{D}(t-t''-t')}\!\Bigg(\!\big[\left(\bar{\mathcal{A}}\!-\!\bar{\mathcal{C}}\!+\!\sqrt{\mathcal{D}}\right)\!N\avg{\delta\mathcal{B}_{j}(t')\delta\mathcal{C}_{j}(t'')}_{e}\!+\!2\bar{\mathcal{B}}\sqrt{N}\avg{\delta\mathcal{C}_{j}(t')\delta\mathcal{C}_{j}(t'')}_{e}\big]\!\frac{\!-\!\bar{\mathcal{A}}\!+\!\bar{\mathcal{C}}\!+\!\sqrt{\mathcal{D}}}{2\sqrt{\mathcal{D}}}\!\Bigg) \nonumber \\
&&+e^{\!-\!\frac{1}{2}\sqrt{D}t}\!\left(\!e^{\frac{1}{2}\sqrt{D}(t'-t'')}\!-\!e^{\frac{1}{2}\sqrt{D}(t'+t'')}\!\right)\!\left(\!\left[\left(\bar{\mathcal{A}}\!-\!\bar{\mathcal{C}}\!-\!\sqrt{\mathcal{D}}\right)\!N\avg{\delta\mathcal{B}_{j}(t')\delta\mathcal{B}_{j}(t'')}_{e}\!+\!2\bar{\mathcal{B}}N\avg{\delta\mathcal{C}_{j}(t')\delta\mathcal{B}_{j}(t'')}_{e}\right]\!\frac{\sqrt{N}\bar{\mathcal{B}}}{\sqrt{\mathcal{D}}}\!\right) \nonumber \\
&&+e^{\frac{1}{2}\sqrt{D}(t'+t''-t)}\!\Bigg(\!\big[\left(\bar{\mathcal{A}}\!-\!\bar{\mathcal{C}}\!-\!\sqrt{\mathcal{D}}\right)\!N\avg{\delta\mathcal{B}_{j}(t')\delta\mathcal{C}_{j}(t'')}_{e}\!+\!2\bar{\mathcal{B}}\sqrt{N}\avg{\delta\mathcal{C}_{j}(t')\delta\mathcal{C}_{j}(t'')}_{e}\big]\!\frac{\bar{\mathcal{A}}\!-\!\bar{\mathcal{C}}\!-\!\sqrt{\mathcal{D}}}{2\sqrt{\mathcal{D}}}\!\Bigg) \nonumber \\
&&+e^{\frac{1}{2}\sqrt{D}(t'-t''-t)}\!\Bigg(\!\Big[\left(\bar{\mathcal{A}}\!-\!\bar{\mathcal{C}}\!-\!\sqrt{\mathcal{D}}\right)\!N\avg{\delta\mathcal{B}_{j}(t')\delta\mathcal{C}_{j}(t'')}_{e}\!+\!2\bar{\mathcal{B}}\sqrt{N}\avg{\delta\mathcal{C}_{j}(t')\delta\mathcal{C}_{j}(t'')}_{e}\Big]\!\frac{\!-\!\bar{\mathcal{A}}\!+\!\bar{\mathcal{C}}\!-\!\sqrt{\mathcal{D}}}{2\sqrt{\mathcal{D}}}\!\Bigg)\Bigg)\Bigg). \label{eq:suppeta2}
\end{eqnarray}
Here we have, once again, neglected the contributions from the fluctuations contained in $\delta\mathcal{A}(t)$ and $\mathcal{B}_{0}$ since they are suppressed by a factor of at least $d\mathcal{F}/N$ compared to the terms above. In deriving Eq.~\eqref{eq:eta2tr}, we have used that $\hat{a}_{\text{cell}}^{(2)}$ consists of sums of the form
\begin{equation} \label{eq:sum1}
\frac{1}{N}\sum_{l=1}^{N-1}\sum_{j=1}^{N}\sum_{j'=1}^{N}e^{-2i\pi/N(j-j')l}\delta X_{j}(t')\delta X_{j'}(t''),
\end{equation}
where $X_{j}$ could e.g. denote $\mathcal{B}_{j}$. For $\eta_{\text{read},2}$, we calculate $\avg{\hat{a}_{\text{cell}}^{(0)}\hat{a}_{\text{cell}}^{(2)}}$ and the average of Eq.~\eqref{eq:sum1} is approximatively $N\avg{\delta X_{j}(t')\delta X_{j}(t'')}_{e}$ because $\avg{\delta X_{j}(t')\delta X_{j'}(t'')}=0$, if $j\neq j'$, since the motion of different atoms are uncorrelated and we have assumed that $N-1\approx N$. All correlations appearing in Eq.~\eqref{eq:eta2tr} are thus single atom correlations and the index $j$ is kept to indicate this. The correlations contained in $\eta_{\text{read},2}$ can be treated analytically, in a similar fashion as the correlations in $\avg{\abs{\theta_{j}(t)}^{2}}$ for the write process, but we have instead simulated the correlations numerically for the previously mentioned Cs-cells.

\subsection{Numerical simulation}

\begin{figure} [H]
\centering
\includegraphics[width=0.4\textwidth]{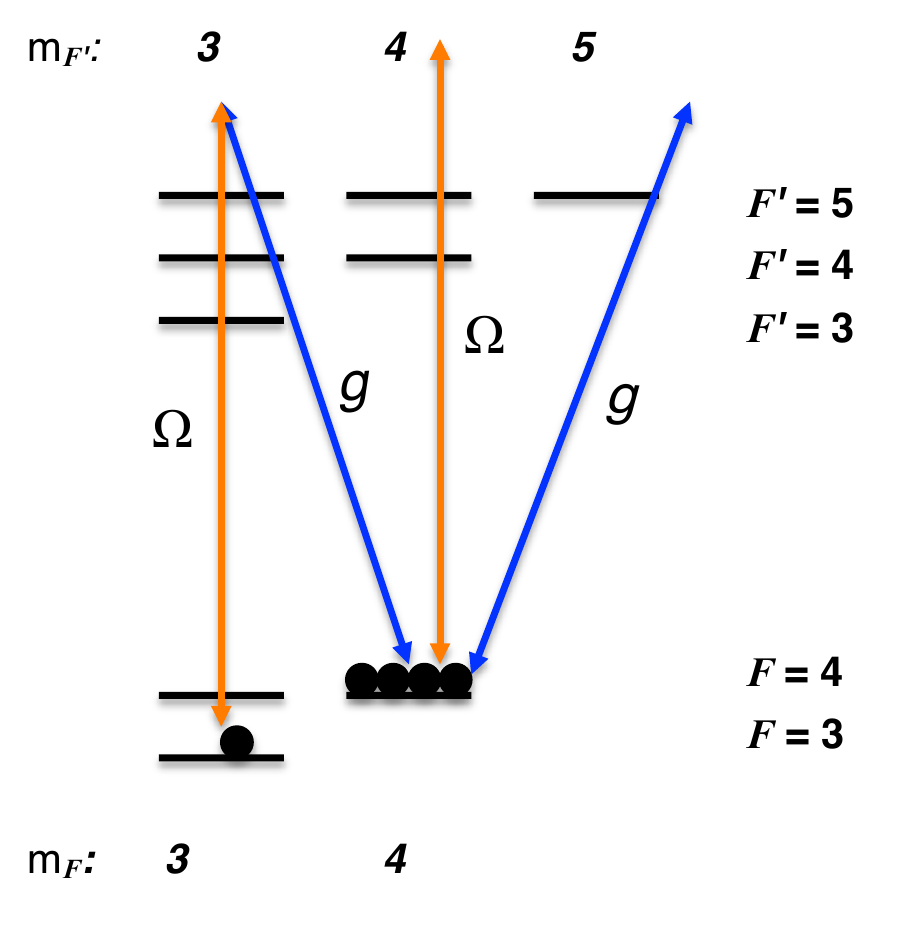}
\caption[Readout realized in ${}^{133}$Cs]{Schematic view of the readout realized in the hyperfine levels of $6^{2}S_{1/2}$ and $6^{2}P_{3/2}$ in ${}^{133}$Cs. We imagine the single exitation to be stored in state $\ket{F=3,m_{F}=3}$ while the macroscopically populated state is $\ket{F=4,m_{F}=4}$. Note that the classical drive also couples $\ket{F=4,m_{F}=4}$ to $\ket{F'=4,m_{F'}=4}$ and $\ket{F'=5,m_{F'}=4}$, which can pump atoms out of $\ket{F=4,m_{F}=4}$. These couplings are however sufficiently suppressed by the large splitting of $2\pi\cdot 9.2$ GHz between the ground states. }
\label{fig:figureS5}
\end{figure}

The simulations are performed in the same way as for the write process. An extra difficulty is, however, that we consider the coupling between the light fields and the extra levels in ${}^{133}$Cs. We assume that the readout process has the level structure shown in \figref{fig:figureS5}.
The couplings to the extra levels result in extra coupling terms in the expressions for $\mathcal{A},\mathcal{B}_{j}$ and $\mathcal{C}_{j}$, which we include, but the expression for $\eta_{\text{read},2}$ is still the same as given in Eq.~\eqref{eq:eta2tr}. Note, however, that a cavity detuning of the quantum field (appearing in the expression for $\mathcal{A}$) on the order of $\kappa_{1}$ is needed to compensate the phases resulting from some of these additional couplings. The starting point of our numerical simulations is therefore Eq.~\eqref{eq:suppeta2}, where we can change the order of integration and introduce the variables $u=t'+t''$ and $s=t'-t''$ since the correlations only depend on the time difference $\abs{t'-t''}$. Performing the integrals over $t$ and $u$ analytically, allows us to write
\begin{eqnarray} \label{eq:suppeta22}
\eta_{\text{read},2}&=&\int_{0}^{\tau_{\text{read}}}(h_{1}(\tau_{\text{read}},s)\avg{\delta\mathcal{B},\delta\mathcal{B}}_{e}(s)+h_{2}(\tau_{\text{read}},s)\avg{\delta\mathcal{B},\delta\mathcal{C}}_{e}(s) \nonumber \\
&&+h_{3}(\tau_{\text{read}},s)\avg{\delta\mathcal{C},\delta\mathcal{B}}_{e}(s)+h_{4}(\tau_{\text{read}},s)\avg{\delta\mathcal{C},\delta\mathcal{C}}_{e}(s))ds,
\end{eqnarray}
where $h_{1}(\tau_{\text{read}},s),h_{2}(\tau_{\text{read}},s),h_{3}(\tau_{\text{read}},s)$ and $h_{4}(\tau_{\text{read}},s)$ are functions of $s$ and $\tau_{\text{read}}$, which are obtained from the integration over $t$ and $u$. We have once again introduced the short notation for the correlations $\avg{\delta\mathcal{B}_{j}(t')\delta\mathcal{C}_{j}(t'')}_{e}=\avg{\delta\mathcal{B},\delta\mathcal{C}}_{e}(s)$. Note that $\avg{\delta\mathcal{B},\delta\mathcal{C}}_{e}(s)\to0$ for $s\to\infty$ similar to the situation in the write process, i.e., the coupling of an atom at time $t$ is uncorrelated from its initial coupling if $t$ is large. We can therefore introduce a cutoff $s_{\text{max}}$ in the integral in Eq.~\eqref{eq:suppeta22} such that we can evaluate $\eta_{\text{read},2}$ for an arbitrary length of the readout pulse $\tau_{\text{read}}$ without additional numerical difficulty. We then numerically evaluate $\eta_{\text{read},2}$ from Eq.~\eqref{eq:suppeta22} by simulating the decay of the correlations similar to the simulation of the write process. From the numerical simulation, we find that the term $\avg{\delta\mathcal{C}_{j}\delta\mathcal{C}_{j}}_{e}$ dominates $\eta_{\text{read},2}$. This term describes loss of the excitation due to spontaneous emission to modes not confined by the cavity.

\section{Errors}

Here we give the detailed expression for the probability ($p_{1}$) to read out incoherent photons to first order.  The incoherent photons are being readout from assymetric modes described by the operators
\begin{equation}
\hat{S}_{l}=\frac{1}{\sqrt{N}}\sum_{j=1}^{N}e^{2i\pi(j-1)l/N}\hat{\sigma}_{01}^{(j)},
\end{equation}
where $l=\{1,2,\ldots,N-1\}$. Note that the symmetric Dicke mode is described by $\hat{S}_{0}$. From the pertubative expansion of $\hat{a}_{\text{cell}}$ we find that the assymetric modes gives a first order contribution of
\begin{eqnarray}
\hat{a}^{(1)}_{\text{cell}}(t)&=&\sqrt{\epsilon}\int_{0}^{t}dt'\frac{e^{\frac{1}{2}(\bar{\mathcal{A}}+\bar{\mathcal{C}})(t-t')}}{2\sqrt{\mathcal{D}}}e^{\bar{\mathcal{C}}t'}\frac{1}{\sqrt{N}}\sum_{l=1}^{N-1}\sum_{j=1}^{N}e^{-2i\pi(j-1)l/N}\Big( \nonumber \\
&& \left(e^{\frac{1}{2}\sqrt{\mathcal{D}}(t-t')}-e^{-\frac{1}{2}\sqrt{\mathcal{D}}(t-t')}\right)\left(\left(\bar{\mathcal{A}}-\bar{\mathcal{C}}\right)\delta\mathcal{B}_{j}(t')+2\bar{\mathcal{B}}\delta\mathcal{C}_{j}(t')\right)\hat{S}_{l} \nonumber \\
&&+ \left(e^{\frac{1}{2}\sqrt{\mathcal{D}}(t-t')}+e^{-\frac{1}{2}\sqrt{\mathcal{D}}(t-t')}\right)\sqrt{\mathcal{D}}\delta\mathcal{B}_{j}(t')\hat{S}_{l}\Big).
\end{eqnarray}
The probability to read out the incoherent photons is then
\begin{equation} \label{eq:pbad}
p_{1}=\frac{\kappa_{2}^{2}\kappa_{1}}{4}\int_{0}^{\tau_{read}}dt\int_{0}^{t}dt'\int_{0}^{t}dt''e^{-\kappa_{2}/2(2t-t'-t'')}\avg{(\hat{a}^{(1)}_{\text{cell}}(t'))^{\dagger}\hat{a}^{(1)}_{\text{cell}}(t'')}.
\end{equation}
We can numerically evaluate the correlations contained in Eq.~\eqref{eq:pbad} in a similar fashion as the correlations in $\eta_{\text{read},2}$, i.e., we simulate the experimental Cs-cells. Thereby we get the results presented in Fig. 3b in the article.

\subsection{DLCZ error analysis}
In order to fully characterize the performance of a DLCZ repeater based on the room temperature cells, we investigate how the errors from incoherent photons propagate in the repeater. First we consider the state being produced in the entanglement generation step (see above). We define the following parameters:
\begin{itemize}
\item $\eta_{\text{write}}$ :  The write efficiency, which basically is the probability to have a Dicke state in the ensemble conditioned on a quantum photon being emitted. If we are not in the right state we are dominated by the intitial state and thus with probability $1-\eta_{\text{write}}$, we assume the atomic state to be $\ket{00\ldots0}$.
\item $p_{e}$ : The excitation probability, which depends on the driving strength.
\item $P_{\text{d}}$ : The dark count probability of a detector. We estimate this as $P_{\text{d}}\sim r_{dark} t_{\text{int}}$ where $r_{dark}$ is the dark count rate and $t_{\text{int}}$ is the length of the driving pulse.
\item $\eta_{\text{d}}$ : The detection efficiency, which is determined by the total efficiency of the detector, the outcoupling losses and the transmission losses from the cavity to the detector.
\end{itemize}

To second order in the excitation probability, the state of the two ensembles following a single click in a detector at the central station is described by the density matrix
\begin{eqnarray} \label{eq:rho1}
\rho_{\text{success}}&=&\Bigg[ \Big[p_{e}^{2}(1-\eta_{\text{write}})^{2}(2\eta_{\text{d}}-2\eta_{\text{d}}^{2})(1-P_{\text{d}})^{2}+2p_{e}^{2}(1-\eta_{\text{write}})^{2}(1-\eta_{\text{d}})^{2}P_{\text{d}}(1-P_{\text{d}})\nonumber \\
&&+2(1-p_{e})^{2}P_{\text{d}}(1-P_{\text{d}})+4p_{e}(1-\eta_{\text{write}})(1-p_{e})(1-\eta_{\text{d}})P_{\text{d}}(1-P_{\text{d}})\nonumber \\
&&+4p_{e}^{2}(1-\eta_{\text{write}})^{2}(\eta_{\text{d}}-\eta_{\text{d}}^2)(1-P_{\text{d}})^{2}+4p_{e}^{2}(1-\eta_{\text{write}})^{2}(1-\eta_{\text{d}})^{2}P_{\text{d}}(1-P_{\text{d}})\Big]\ket{\mathbf{00}}\bra{\mathbf{00}} \nonumber \\
&&+\Big[p_{e}^{2}\eta_{\text{write}}^{2}(2\eta_{\text{d}}-2\eta_{\text{d}}^{2})(1-P_{\text{d}})^{2}+2p_{e}^{2}\eta_{\text{write}}^{2}(1-\eta_{\text{d}})^{2}P_{\text{d}}(1-P_{\text{d}})\Big]\ket{\mathbf{11}}\bra{\mathbf{11}} \nonumber \\
&&+\Big[p_{e}^{2}\eta_{\text{write}}(1-\eta_{\text{write}})(2\eta_{\text{d}}-2\eta_{\text{d}}^{2})(1-P_{\text{d}})^{2}+2p_{e}^{2}\eta_{\text{write}}(1-\eta_{\text{write}})(1-\eta_{\text{d}})^{2}P_{\text{d}}(1-P_{\text{d}}) \nonumber \\
&&+2p_{e}\eta_{\text{write}}(1-p_{e})(1-\eta_{\text{d}})P_{\text{d}}(1-P_{\text{d}})+4p_{e}^{2}\eta_{\text{write}}(1-\eta_{\text{write}})(\eta_{\text{d}}-\eta_{\text{d}}^{2})(1-P_{\text{d}})^{2}\nonumber \\
&&+4p_{e}^{2}\eta_{\text{write}}(1-\eta_{\text{write}})(1-\eta_{\text{d}})^{2}P_{\text{d}}(1-P_{\text{d}})\Big](\ket{\mathbf{01}}\bra{\mathbf{01}}+\ket{\mathbf{10}}\bra{\mathbf{10}}) \nonumber \\
&&+\Big[2p_{e}^{2}\eta_{\text{write}}^{2}(\eta_{\text{d}}-\eta_{\text{d}}^{2})(1-P_{\text{d}})^{2}+2p_{e}^{2}\eta_{\text{write}}^{2}(1-\eta_{\text{d}})^{2}P_{\text{d}}(1-P_{\text{d}})\Big](\ket{\mathbf{20}}\bra{\mathbf{20}}+\ket{\mathbf{02}}\bra{\mathbf{02}}) \nonumber \\
&&+2p_{e}\eta_{\text{write}}(1-p_{e})\eta_{\text{d}}(1-P_{\text{d}})^{2}\ket{\mathbf{\Psi}}\bra{\mathbf{\Psi}} \Bigg]\frac{1}{\mathcal{N}},
\end{eqnarray}
where we have defined $\ket{\mathbf{0}}=\ket{00\ldots 0}$, $\ket{\mathbf{1}}=\ket{\text{Dicke}}$, $\ket{\mathbf{2}}=\frac{1}{\sqrt{N}}\sum_{i,j=1}^{N}\ket{1}_{i}\ket{1}_{j}\bra{0}_{i}\bra{0}_{j}\ket{00\ldots0}$ and $\ket{\mathbf{\Psi}}=\frac{1}{\sqrt{2}}(\ket{\mathbf{01}}+\ket{\mathbf{10}})$. $\mathcal{N}=P_{\text{success}}$ is a normalization constant, which gives the success probability of the operation. Note that we have assumed number resolving detectors. It is seen from Eq.~\eqref{eq:rho1} that we can write $\rho_{\text{success}}=a_{0}\ket{\mathbf{\Psi}}\bra{\mathbf{\Psi}}+b_{0}\ket{\mathbf{00}}\bra{\mathbf{00}}+c_{0}(\ket{\mathbf{01}}\bra{\mathbf{01}}+\ket{\mathbf{10}}\bra{\mathbf{10}})+d_{0}\ket{\mathbf{11}}\bra{\mathbf{11}}+e_{0}(\ket{\mathbf{02}}\bra{\mathbf{02}}+\ket{\mathbf{20}}\bra{\mathbf{20}})$.

We now consider the entanglement swapping of two states of the form $\rho_{\text{success}}$ using the setup shown in Fig.~\ref{fig:figure1}. In the swap, an ensemble from each entangled pair is read out and the corresponding photons are combined on a balanced beam splitter. With probability $a_{0}^{2}$, we are swapping two states of the form $\ket{\mathbf{\Psi}}$ and the state after a successful swap is
\begin{eqnarray} \label{eq:rho2}
\rho_{a_{0}^{2}} &=&\Bigg[\frac{1}{4}\Big[2\eta_{\text{dr}}(1-\eta_{\text{dr}})(1-P_{\text{dr}})^{2}+2(1-\eta_{\text{dr}})^{2}P_{\text{dr}}(1-P_{\text{dr}})\Big]\ket{\mathbf{00}}\bra{\mathbf{00}} \nonumber \\
&&+\frac{1}{4}\Big[2P_{\text{dr}}(1-P_{\text{dr}})\Big]\ket{\mathbf{11}}\bra{\mathbf{11}} \nonumber \\
&&+\frac{1}{4}\Big[2(1-\eta_{\text{dr}})P_{\text{dr}}(1-P_{\text{dr}})\Big](\ket{\mathbf{01}}\bra{\mathbf{01}}+\ket{\mathbf{10}}\bra{\mathbf{10}}) \nonumber \\
&&+\frac{1}{2}\eta_{\text{dr}}(1-P_{\text{dr}})^{2}\ket{\Psi}\bra{\Psi}\Bigg]\frac{1}{\mathcal{N}'},
\end{eqnarray}
where we have introduced the total readout detection efficiency $\eta_{\text{dr}}$, which is determined by the readout efficiency, the outcoupling losses and the efficiency of the detectors.  In contrast to the detection efficiency in the entanglement generation, the readout detection efficiency does not include fiber losses since the entanglement swap is a local process. $\mathcal{N'}$ is a normalization constant. Furthermore, we have defined the readout dark count rate $P_{\text{dr}}$, which contains the probability of reading out incoherent photons from the ensembles and the detector dark counts, i.e. $P_{\text{dr}}=r_{\text{dark}}\tau_{\text{read}}+\eta_{\text{dr}}p_{1}$. Here $r_{\text{dark}}\tau_{\text{read}}$ is the dark count rate of the detectors and $p_{1}$ is the probability to emit incoherent photons. As previously described $p_{1}$ is mainly determined by the inefficiency of the optical pumping in the initialization of the ensembles and the memory time of the ensemble. We will neglect the limited memory time and simply assume a fixed value of $p_{1}$ from the inefficiency of the optical pumping. From Eq. \eqref{eq:rho2}, we can express $\rho_{a_{0}^{2}}$ as
\begin{eqnarray}
\rho_{a_{0}^{2}}&=&\frac{1}{\mathcal{N}'}\Bigg[\frac{1}{4}\alpha\ket{\mathbf{00}}\bra{\mathbf{00}}+\frac{1}{4}(\beta-\eta_{\text{dr}}(1-P_{\text{dr}})^{2})(\ket{\mathbf{01}}\bra{\mathbf{01}}+\ket{\mathbf{10}}\bra{\mathbf{10}}) \nonumber \\
&&+\frac{1}{4}\lambda\ket{\mathbf{11}}\bra{\mathbf{11}}+\frac{1}{2}\eta_{\text{dr}}(1-P_{\text{dr}})^{2}\ket{\mathbf{\Psi}}\bra{\mathbf{\Psi}}\Bigg],\qquad
\end{eqnarray}
with $\alpha,\beta$, and $\lambda$ given by Eq.~\eqref{eq:rho2}. Considering all the combinations from swapping  two states of the form $\rho_{\text{success}}$, we find that the output state can be written as
\begin{eqnarray} \label{eq:rho3}
\rho_{\text{swap},1}&=&\frac{1}{\mathcal{N''}}\Bigg[\Big[\frac{a_{0}^{2}}{4}\alpha+b_{0}^{2}\lambda+c_{0}^{2}\alpha+b_{0}a_{0}\beta+c_{0}a_{0}\alpha+2b_{0}c_{0}\beta+2b_{0}e_{0}\tilde{\beta}+2c_{0}e_{0}\tilde{\alpha}+e_{0}^{2}\tilde{\gamma}+a_{0}e_{0}\tilde{\alpha}\Big]\ket{\mathbf{00}}\bra{\mathbf{00}} \qquad \nonumber \\
&&+\Big[\frac{a_{0}^{2}}{4}(\beta-\eta_{\text{dr}}(1-P_{\text{dr}})^{2})+\frac{a_{0}b_{0}}{2}\lambda+a_{0}c_{0}\beta+\frac{a_{0}d_{0}}{2}\alpha+b_{0}c_{0}\lambda+b_{0}d_{0}\beta+c_{0}^{2}\beta+c_{0}d_{0}\alpha \nonumber \\
&&+\frac{a_{0}e_{0}}{2}\tilde{\beta}+d_{0}e_{0}\tilde{\alpha}\Big](\ket{\mathbf{01}}\bra{\mathbf{01}}+\ket{\mathbf{10}}\bra{\mathbf{10}}) \nonumber \\
&&+\Big[\frac{a_{0}^{2}}{4}\lambda+c_{0}^{2}\lambda+d_{0}^{2}\alpha+c_{0}a_{0}\lambda+d_{0}a_{0}\beta+2c_{0}d_{0}\beta\Big]\ket{\mathbf{11}}\bra{\mathbf{11}}\nonumber \\
&&+\Big[\frac{a_{0}e_{0}}{2}\beta+b_{0}e_{0}\lambda+c_{0}e_{0}\beta+2e_{0}^{2}\tilde{\beta}\Big](\ket{\mathbf{02}}\bra{\mathbf{02}}+\ket{\mathbf{20}}\bra{\mathbf{20}}) \nonumber \\
&&+\Big[\frac{a_{0}e_{0}}{2}\lambda+c_{0}e_{0}\lambda+d_{0}e_{0}\beta\Big](\ket{\mathbf{12}}\bra{\mathbf{12}}+\ket{\mathbf{21}}\bra{\mathbf{21}}) \nonumber \\
&&+e_{0}^{2}\lambda\ket{\mathbf{22}}\bra{\mathbf{22}} \nonumber \\
&&+\frac{a_{0}^{2}}{2}\eta_{\text{dr}}(1-P_{\text{dr}})\ket{\mathbf{\Psi}}\bra{\mathbf{\Psi}}\Bigg],
\end{eqnarray}
where we have defined
\begin{eqnarray}
\tilde{\alpha}&=&3\eta_{\text{dr}}(1-\eta_{\text{dr}})^{2}(1-P_{\text{dr}})^{2}+2(1-\eta_{\text{dr}})^{3}P_{\text{dr}}(1-P_{\text{dr}})\\
\tilde{\beta}&=&2\eta_{\text{dr}}(1-\eta_{\text{dr}})(1-P_{\text{dr}})^{2}+2(1-\eta_{\text{dr}})^{2}P_{\text{dr}}(1-P_{\text{dr}}) \\
\tilde{\gamma}&=&4\eta_{\text{dr}}(1-\eta_{\text{dr}})^{3}(1-P_{\text{dr}})^{2}+2(1-\eta_{\text{dr}})^{4}P_{\text{dr}}(1-P_{\text{dr}}),
\end{eqnarray}
and $\mathcal{N}''=P_{\text{swap},1}$ is a normalization constant, which gives the success probability of the swap operation. It is seen that we can write
\begin{eqnarray}
\rho_{\text{swap},1}&=&a_{1}\ket{\mathbf{\Psi}}\bra{\mathbf{\Psi}}+b_{1}\ket{\mathbf{00}}\bra{\mathbf{00}}+c_{1}(\ket{\mathbf{01}}\bra{\mathbf{01}}+\ket{\mathbf{10}}\bra{\mathbf{10}})+d_{1}\ket{\mathbf{11}}\bra{\mathbf{11}}+e_{1}(\ket{\mathbf{02}}\bra{\mathbf{02}}+\ket{\mathbf{20}}\bra{\mathbf{20}})\nonumber \\
&&+f_{1}(\ket{\mathbf{12}}\bra{\mathbf{12}}+\ket{\mathbf{21}}\bra{\mathbf{21}})+g_{1}\ket{\mathbf{22}}\bra{\mathbf{22}}
\end{eqnarray}
with constants $a_{1},b_{1},c_{1},d_{1},e_{1},f_{1}$ and $g_{1}$ determined by Eq.~\eqref{eq:rho3}.

The vacuum part of the swapped state grows exponentially with the number of swaps in the DLCZ protocol \cite{sangouardrev} and it is therefore necessary to perform a final postselection  where two entangled states are combined (see \figref{fig:figure1}).
\begin{figure}
\centering
\includegraphics[scale=1]{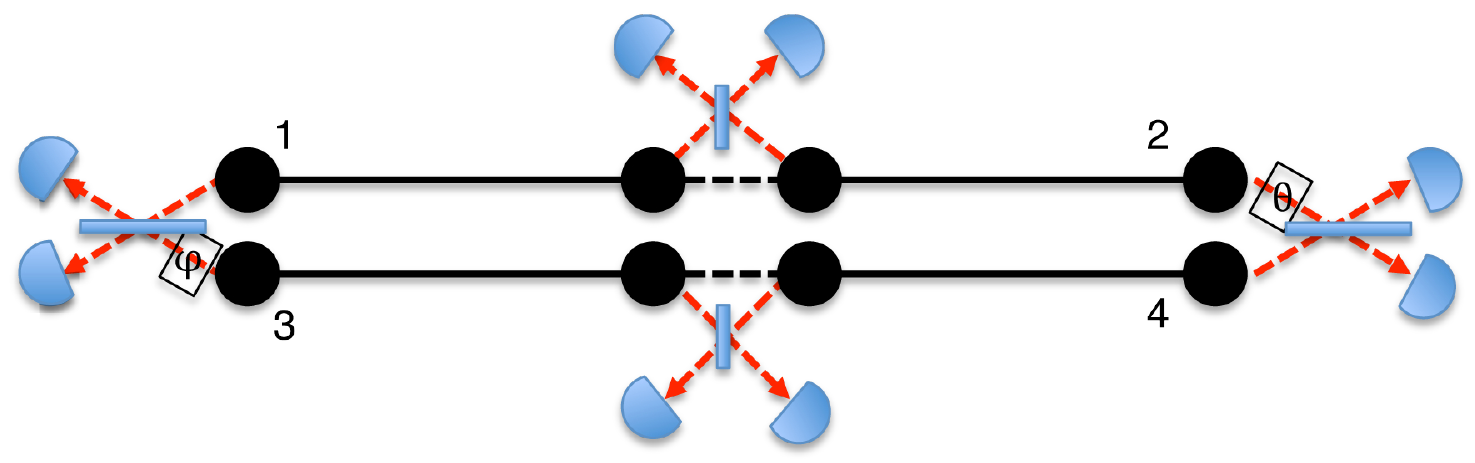}
\caption{Sketch of the postselection procedure after a single step of entanglement swapping (middle stations). The entanglement swaps creates entanglement between ensembles 1,2 and 3,4, respectively.  Ensembles 1,2,3 and 4 are then read out and we condition on a single click at both stations. The phases $\phi$ and $\theta$ are ideally equal. }.
\label{fig:figure1}
\end{figure}
We assume two parties named Alice and Bob who share two entangled pairs such that Alice has ensemble 1 and 3 and Bob has ensemble 2 and 4. Ensemble 1 and 2 are entangled and so is ensemble 3 and 4. The intial state is then
 \begin{eqnarray}
\rho_{\text{intial}}&=&\Big(a\ket{\mathbf{\Psi}}\bra{\mathbf{\Psi}}+b\ket{\mathbf{00}}\bra{\mathbf{00}}+c(\ket{\mathbf{01}}\bra{\mathbf{01}}+\ket{\mathbf{10}}\bra{\mathbf{10}})+d\ket{\mathbf{11}}\bra{\mathbf{11}} \nonumber \\
&&+e(\ket{\mathbf{02}}\bra{\mathbf{02}}+\ket{\mathbf{20}}\bra{\mathbf{20}})+f(\ket{\mathbf{12}}\bra{\mathbf{12}}+\ket{\mathbf{21}}\bra{\mathbf{21}})+g\ket{\mathbf{22}}\bra{\mathbf{22}}\Big)_{1,2} \nonumber \\
&& \otimes\Big(a\ket{\mathbf{\Psi}}\bra{\mathbf{\Psi}}+b\ket{\mathbf{00}}\bra{\mathbf{00}}+c(\ket{\mathbf{01}}\bra{\mathbf{01}}+\ket{\mathbf{10}}\bra{\mathbf{10}})+d\ket{\mathbf{11}}\bra{\mathbf{11}} \nonumber \\
&&+e(\ket{\mathbf{02}}\bra{\mathbf{02}}+\ket{\mathbf{20}}\bra{\mathbf{20}})+f(\ket{\mathbf{12}}\bra{\mathbf{12}}+\ket{\mathbf{21}}\bra{\mathbf{21}})+g\ket{\mathbf{22}}\bra{\mathbf{22}}\Big)_{3,4}
\end{eqnarray}
All ensembles are now readout in, e.g., a cryptography scheme and a success is conditioned on both Alice and Bob recording a single click. The total success probability is
\begin{eqnarray}
P_{\text{ps}}&=&\frac{a^{2}}{2}(\alpha_{\text{ps}}\beta_{\text{ps}}+\gamma_{\text{ps}}^{2})+b^{2}\alpha_{\text{ps}}^{2}+2c^{2}\alpha_{\text{ps}}\beta_{\text{ps}}+d^{2}\beta_{\text{ps}}^{2}+2c^{2}\gamma_{\text{ps}}^{2}+2ab\alpha_{\text{ps}}\gamma_{\text{ps}}\nonumber \\
&&+2ac(\alpha_{\text{ps}}\beta_{\text{ps}}+\gamma_{\text{ps}}^{2})+2ad\beta_{\text{ps}}\gamma_{\text{ps}}+4bc\alpha_{\text{ps}}\gamma_{\text{ps}}+4cd\beta_{\text{ps}}\gamma_{\text{ps}}+2bd\gamma_{\text{ps}}^{2} \nonumber \\
&&+2ae(\alpha_{\text{ps}}\tilde{\beta}_{\text{ps}}+\gamma_{\text{ps}}\beta_{\text{ps}})+2af(\gamma_{\text{ps}}\tilde{\beta}_{\text{ps}}+\beta_{\text{ps}}^{2})+2ag\beta_{\text{ps}}\tilde{\beta}_{\text{ps}}+4be\alpha_{\text{ps}}\beta_{\text{ps}} \nonumber \\
&&+4bf\gamma_{\text{ps}}\beta_{\text{ps}}+2bg\beta_{\text{ps}}^{2}+4ce(\gamma_{\text{ps}}\beta_{\text{ps}}+\alpha_{\text{ps}}\tilde{\beta}_{\text{ps}})+4cf(\beta_{\text{ps}}^{2}+\gamma_{\text{ps}}\tilde{\beta}_{\text{ps}})+4cg\beta_{\text{ps}}\tilde{\beta}_{\text{ps}}+4de\gamma_{\text{ps}}\tilde{\beta}_{\text{ps}} \nonumber \\
&&+4df\beta_{\text{ps}}\tilde{\beta}_{\text{ps}}+2dg\tilde{\beta}_{\text{ps}}^{2}+2e^2(\alpha_{\text{ps}}\tilde{\gamma}_{\text{ps}}+\beta_{\text{ps}}^{2})+4eg\beta_{\text{ps}}\tilde{\gamma}_{\text{ps}}+4fg\tilde{\beta}_{\text{ps}}\tilde{\gamma}_{\text{ps}}+g^{2}\tilde{\gamma}_{\text{ps}}^{2} \nonumber \\
&&+4fe(\gamma_{\text{ps}}\tilde{\gamma}_{\text{ps}}+\tilde{\beta}_{\text{ps}}\beta_{\text{ps}})+2f^{2}(\tilde{\gamma}_{\text{ps}}\beta_{\text{ps}}+\tilde{\beta}^{2}),
\end{eqnarray}
where we have defined
\begin{eqnarray}
\alpha_{\text{ps}}&=&2P_{\text{dr}}(1-P_{\text{dr}}) \\
\beta_{\text{ps}}&=&2\eta_{\text{dr}}(1-\eta_{\text{dr}})(1-P_{\text{dr}})^{2}+2(1-\eta_{\text{dr}})^{2}P_{\text{dr}}(1-P_{\text{dr}}) \\
\gamma_{\text{ps}}&=&\eta_{\text{dr}}(1-P_{\text{dr}})^{2}+2(1-\eta_{\text{dr}})P_{\text{dr}}(1-P_{\text{dr}}) \\
\tilde{\beta}_{\text{ps}}&=&3\eta_{\text{dr}}(1-\eta_{\text{dr}})^{2}(1-P_{\text{dr}})^{2}+2(1-\eta_{\text{dr}})^{3}P_{\text{dr}}(1-P_{\text{dr}}) \\
\tilde{\gamma}_{\text{ps}}&=&4\eta_{\text{dr}}(1-\eta_{\text{dr}})^{3}(1-P_{\text{dr}})^{2}+2(1-\eta_{\text{dr}})^{4}P_{\text{dr}}(1-P_{\text{dr}}).
\end{eqnarray}
The postselected fidelity of the state is
\begin{equation}
F_{\text{ps}}=\frac{\frac{a^{2}}{4}\eta_{\text{read}}^{2}\eta_{\text{dr}}^{2}(1-P_{\text{dr}})^{4}(1+\cos(\phi-\theta))+\frac{a^{2}}{4}(\gamma_{\text{ps}}^{2}-\eta_{\text{read}}^{2}\eta_{\text{dr}}^{2}(1-P_{\text{dr}})^{4})+c^{2}\gamma_{\text{ps}}^{2}+ac\gamma_{\text{ps}}^{2}}{P_{\text{ps}}},
\end{equation}
where the phases $\phi$, $\theta$ can be different due to variations in the path lengths of the photons being read out. In the ideal case, we have $\phi=\theta$ for which the fidelity is maximal.

The rate of a DLCZ repeater based on the room temperature cells can be estimated as \cite{sangouardrev}
\begin{equation}
r\approx\left(\frac{2}{3}\right)^{n+1}P_{0}P_{\text{swap},1}P_{\text{swap},2}\ldots P_{\text{swap},n}P_{\text{ps}}\frac{2^{n}c}{L}
\end{equation}
where $n$ is the number of swap levels in the repeater and $L$ is the distance. Note, however, that the scalability of the room temperature cells enables spatially multiplexing, which both increases the rate of the repeater and decreases the necessary memory time of the ensembles \cite{sangouardrev}. Assuming that $2M$ ensembles are used at each repeater station, the rate will increase by a factor of $M$.
We have considered a basic repeater segment consisting of only a single swap without multiplexing in order to estimate the rate and fidelity of a distributed pair. We assume that the distance to distribute entanglement over is $L=80$ km and that the losses in the fibers are given by the attenuation length at telecom wavelengths, which is $\sim$20 km. Furthermore, we assume SPD efficiencies of 95\% and dark count rates of 1 Hz. This reflects what is possible with current superconducting detectors \cite{saewoo, smith}. We then perform an optimization in all the parameters characterizing the cells, e.g. the excitation probability, the write time and the readout time. We include experimental imperfections such as outcoupling losses of around $10\%$ and intracavity losses of $2\%$. As a result we find that a pair with fidelity $\sim80\%$ can be distributed with a rate of $\sim0.2$ Hz. We have assumed that $\epsilon\approx0.5\%$ and have neglected effects from finite memory time of the atoms. Furthermore, we have assumed that $\phi=\theta$ (see \figref{fig:figure1}).

\end{document}